\newif\ifsubmode
\submodefalse

\newif\ifprintfig
\printfigtrue

\newif\ifemulate
\emulatetrue

\ifsubmode
  \documentclass[12pt,preprint]{aastex}
  \received{}
  \accepted{}
  \journalid{}{}
  \articleid{}{}
\else
   \documentclass{emulateapj}
   \submitted{{\it To be submitted for publication in ApJ}}
\fi

\usepackage{amsmath}
\usepackage{appendix}

\newcommand{\kms}{\,km~s$^{-1}$}
\newcommand{\Msun}{\mbox{\,$M_{\odot}$}}

\def\lesssim{\mathrel{\hbox{\rlap{\hbox{\lower4pt\hbox{$\sim$}}}\hbox{$<$}}}}
\def\gtrsim{\mathrel{\hbox{\rlap{\hbox{\lower4pt\hbox{$\sim$}}}\hbox{$>$}}}}
\def\spose#1{\hbox to 0pt{#1\hss}}
\def\simlt{\mathrel{\spose{\lower 3pt\hbox{$\mathchar"218$}}
     \raise 2.0pt\hbox{$\mathchar"13C$}}}
\def\simgt{\mathrel{\spose{\lower 3pt\hbox{$\mathchar"218$}}
     \raise 2.0pt\hbox{$\mathchar"13E$}}}

\slugcomment{Draft Version \today}

\shorttitle{Diffuse Ionized Gas in M83}
\shortauthors{Boettcher, Gallagher, and Zweibel}

\begin{document}

\title{Detection of Extraplanar Diffuse Ionized Gas in M83$^{*}$}
\thanks{$^{*}$Based on observations made with the Southern African Large
  Telescope (SALT) under program 2015-2-SCI-004 (PI: E. Boettcher).}

\author{Erin Boettcher$^{1}$, J. S. Gallagher III$^{1}$, Ellen
  G. Zweibel$^{1,2}$}

\affiliation{$^{1}$Department of Astronomy, University of Wisconsin - Madison,
  475 North Charter Street, Madison, WI 53706, USA; \texttt{boettche@astro.wisc.edu}} 
\affiliation{$^{2}$Department of Physics,
  University of Wisconsin - Madison, 475 North Charter Street, Madison, WI
  53706, USA} 


\ifsubmode\else
  \ifemulate\else
     \clearpage
  \fi
\fi


\ifsubmode\else
  \ifemulate\else
     \baselineskip=14pt
  \fi
\fi

\begin{abstract} 

  We present the first kinematic study of extraplanar diffuse ionized gas
  (eDIG) in the nearby, face-on disk galaxy M83 using optical emission-line
  spectroscopy from the Robert Stobie Spectrograph on the Southern African
  Large Telescope. We use a Markov Chain Monte Carlo method to decompose the
  [NII]$\lambda\lambda$6548, 6583, H$\alpha$, and [SII]$\lambda\lambda$6717,
  6731 emission lines into HII region and diffuse ionized gas
  emission. Extraplanar, diffuse gas is distinguished by its emission-line
  ratios ([NII]$\lambda$6583/H$\alpha \gtrsim 1.0$) and its rotational
  velocity lag with respect to the disk ($\Delta v = -24$ \kms in
  projection). With interesting implications for isotropy, the velocity
  dispersion of the diffuse gas, $\sigma = 96$ \kms, is a factor of a few
  higher in M83 than in the Milky Way and nearby, edge-on disk galaxies.  The
  turbulent pressure gradient is sufficient to support the eDIG layer in
  dynamical equilibrium at an electron scale height of $h_{z} = 1$
  kpc. However, this dynamical equilibrium model must be finely tuned to
  reproduce the rotational velocity lag. There is evidence of local bulk flows
  near star-forming regions in the disk, suggesting that the dynamical state
  of the gas may be intermediate between a dynamical equilibrium and a
  galactic fountain flow. As one of the first efforts to study eDIG kinematics
  in a face-on galaxy, this study demonstrates the feasibility of
  characterizing the radial distribution, bulk velocities, and vertical
  velocity dispersions in low-inclination systems.
  
\end{abstract}


\section{Introduction}\label{sec_intro}

The formation and evolution of multi-phase, gaseous galactic halos is affected
by star-formation feedback in galactic disks, determining the pressure in the
midplane, the enrichment of the intergalactic medium, and the distribution of
baryons in the universe. To understand the cycling of mass and energy between
the disk and the halo, we must understand the nature of the disk-halo
interface, including its structure, energetics, and dynamics.

The dynamical state of the warm ionized phase of this interface, known as the
Reynolds Layer in the Milky Way, is not well understood. These extraplanar
diffuse ionized gas (eDIG) layers are common in star-forming disk galaxies
\citep{Lehnert1995, Rossa2003a}, but their observed exponential electron scale
heights tend to greatly exceed their thermal scale heights
\citep[e.g.,][]{Rand1997, Haffner1999, Collins2001}. Thus, it is not known
whether these layers are in dynamical equilibrium, or are evidence of a
non-equilibrium state such as a galactic fountain, a galactic wind, or an
accretion flow.

Studies of the dynamical state of extraplanar gas layers have largely focused
on nearby edge-on disk galaxies \citep[e.g.,][hereafter B16]{Collins2002,
  Barnabe2006, Fraternali2006, Boettcher2016}. In these systems, the gas
density, rotational velocity, and horizontal velocity dispersion can be
determined as functions of height above the disk, and the contributions of
thermal and non-thermal pressure gradients to the vertical structure and
support can be quantified. The observed exponential electron scale heights of
eDIG layers are on the order of $h_{z} = 1$ kpc, and may exceed this value by
a factor of a few \citep[e.g.,][]{Rand1997, Haffner1999, Collins2001}. The
thermal velocity dispersion of a $T \sim 10^{4}$ K gas is $\sigma \sim 10$
\kms, corresponding to a thermal scale height of only $h_{z} = 100 - 200$
pc. The turbulent velocity dispersions - measured parallel to the disk - tend
to be a few tens of \kms, increasing the scale height to only a few hundred
parsecs (\citealt{Heald2006a, Heald2006b}; B16). Even accounting for thermal,
turbulent, magnetic field, and cosmic ray pressure gradients, the eDIG layer
in NGC 891 cannot be supported at $h_{z} = 1$ kpc at $R \le 8$ kpc (B16).

Studies of eDIG layers in low-inclination disk galaxies provide a
complementary - and largely unexploited - perspective on studies of
high-inclination systems. In a face-on galaxy, the vertical velocity
dispersion can be directly measured, and one does not have to rely on the
assumption that the velocity dispersion is isotropic. Additionally, any
vertical bulk flows can be detected, and the radial distribution of the gas
can be determined, with particular attention to the relationship between eDIG
properties and underlying disk features. Fraternali et al. (2004) demonstrated
the feasibility of detecting lagging, extraplanar emission in the
moderately-inclined galaxy NGC 2403 using optical emission-line
spectroscopy. However, this approach has not been widely applied.

By studying a sample of galaxies with a range of inclination angles, the
complete kinematics of eDIG layers can be pieced together. From such a sample,
we can address the three-dimensional density distribution, velocity profile,
and velocity dispersion, and assess any dependence on the underlying disk
properties. Beyond the dynamical state of the eDIG layer itself, these studies
will shed light on the isotropy of the velocity dispersion, turbulent and bulk
motions relevant to magnetic dynamos, and the relationship between the
extraplanar cold, warm, and hot phases.

Here, we perform a study of the eDIG layer in the well-studied disk galaxy M83
(NGC 5236) using optical emission-line spectroscopy from the Robert Stobie
Spectrograph on the Southern African Large Telescope. M83 is a nearby ($D =
4.8$ Mpc; $1" = 23.3$ pc; \citealt{Karachentsev2007}), face-on ($i =
24^{\circ}$; \citealt{Park2001}) spiral galaxy with an SAB(s)c classification
in the Third Reference Catalog of Bright Galaxies
\citep{deVaucouleurs1991}. It has a modest star-formation rate of $SFR = 3.2$
M$_{\odot}$ yr$^{-1}$ \citep{Jarrett2013}, with much of its star formation
concentrated in nuclear star clusters. The mass, morphology, and
star-formation rate of this system are similar to that of the well-studied,
edge-on galaxy NGC 891; thus, a comparison of the eDIG properties in these
galaxies is of particular interest.

Observations of M83 from the radio to the X-ray regimes have revealed a
complex, multi-phase gaseous halo. M83 has an extended HI disk that is
detected to $R > 50$ kpc, with warped and filamentary structure suggestive of
interactions with a companion (Miller et al. 2009, Heald et al. 2016). In the
inner disk, Miller et al. (2009) detect extraplanar HI with a rotational
velocity lag of 40 - 50 \kms in projection and a line-of-sight velocity
dispersion of $\sigma = 10 - 15$ \kms. $\sigma$ may be underestimated if the
wings of the thick disk emission are compromised by the removal of the thin
disk. Within $R = 8$ kpc, there are $5.6 \times 10^{7} \Msun$ of extraplanar
HI, with a comparable amount of mass in high-velocity, neutral clouds. Miller
et al. (2009) interpret these observations as indicative of a galactic
fountain coupled with tidal interactions.

\textit{Chandra X-ray Observatory} observations reveal diffuse, soft X-ray
emission that traces the nucleus and spiral arms of M83 \citep{Soria2002,
  Soria2003, Long2014}. In the starburst nucleus, the diffuse, hot gas has a
temperature of $T \sim 7 \times 10^{6}$ K, a redshifted velocity of $\sim
7000$ \kms, and abundances consistent with enrichment by Type-II supernovae
and stellar winds from Wolf-Rayet stars \citep{Soria2002}. This is suggestive
of diffuse, hot gas near areas of star-formation activity, and perhaps of a
star-formation-driven nuclear outflow. The role of hot halos in the vertical
support of the warm phase and the interaction between galactic outflows and
eDIG layers remain open questions.

The layout of this paper is as follows. In \S2, we discuss the data
acquisition, data reduction, and flux calibration. We detail the use of a
Markov Chain Monte Carlo method to model the emission-line spectra as
superpositions of HII region and diffuse emission in \S3. In \S4.1, we
identify the diffuse emission from its emission-line ratios. In \S4.2, we
discuss the kinematics of the diffuse gas, including the line-of-sight
velocity and velocity dispersion, and we kinematically identify the eDIG
layer. We consider the proximity of our eDIG detections to star-forming
regions in \S4.3, and we estimate the total mass of the layer in \S5. We test
a dynamical equilibrium model of the eDIG layer in \S6. In \S7, we compare our
results to observations of M83, the Milky Way, and nearby edge-on disk
galaxies in the literature, and discuss the merits of both equilibrium and
non-equilibrium models. We summarize and conclude in \S8.

\section{Observations}\label{sec_obs}

We obtained observations on 2016 April 04 - 15 using the Robert Stobie
Spectrograph \citep{Burgh2003, Kobulnicky2003} on the Southern African Large
Telescope (SALT; \citealt{Buckley2006}). The $8'$ longslits are centered on
the nucleus at two position angles, and, where possible, lie along dust lanes,
between spiral arms, and away from HII regions to favor faint extraplanar
emission. We obtained observations at low and moderate spectral
resolution. The former use a $1.5"$ slit and the pg0900 grating at an angle of
$13.625^{\circ}$. This yields wavelength coverage from 3600 \AA $ \ - \ $ 6700
\AA, a dispersion of 0.97 \AA/pixel, and spectral resolution $\texttt{R} =
1100$ ($\sigma = 116$ \kms) at H$\alpha$. Using a $1"$ slit and the pg2300
grating at an angle of $50.0^{\circ}$ provides wavelength coverage from 6200
\AA $ \ - \ $ 7000 \AA, a dispersion of 0.25 \AA/pixel, and spectral
resolution $\texttt{R} = 5490$ ($\sigma = 23$ \kms) at H$\alpha$. Using $2
\times 2$ binning, the spatial plate scale is $0.25"$/pixel. The locations of
the slits are shown in Figure 1, and the coordinates, position angles, and
exposure times are given in Table 1.

\begin{deluxetable*}{cccccc}[t]
\tabletypesize{\scriptsize}
\tablecolumns{6}
\tablewidth{0pt}
\tablecaption{M83 Observing Summary}
\tablehead{ 
\colhead{Slit} &
\colhead{R.A. \tablenotemark{a}} &
\colhead{Decl. \tablenotemark{a}} &
\colhead{P.A. \tablenotemark{b}} &
\colhead{$t_{exp}$ (pg0900) \tablenotemark{c}} &
\colhead{$t_{exp}$ (pg2300) \tablenotemark{d}}\\
\colhead{Label} &
\colhead{(J2000)} &
\colhead{(J2000)} &
\colhead{(deg)} &
\colhead{(s)} &
\colhead{(s)}
}
\startdata
s1 & 13 37 01.1 & -29 51 38 & 4.0 & $2 \times 950$ & $4 \times 850$ \\
s2 & 13 37 00.4 & -29 52 02 & 50.0 & $2 \times 950$ & $6 \times 850$
\enddata
\tablenotetext{a}{The R.A. and Decl. at the center of the slit. R.A. is
  measured in hours, minutes, and seconds; Decl. is measured in degrees,
  arcminutes, and arcseconds.}
\tablenotetext{b}{The position angle measured from north to east.}
\tablenotetext{c}{The exposure time at low spectral resolution.}
\tablenotetext{d}{The exposure time at moderate spectral resolution.}
\end{deluxetable*}

We used the SALT science pipeline\footnote[1]{\url{http://pysalt.salt.ac.za/}}
to perform the initial data reduction, including bias, gain, and cross-talk
corrections and image preparation and mosaicking \citep{Crawford2010}. We then
used the IRAF\footnote[2]{IRAF is distributed by the National Optical
  Astronomy Observatories, which are operated by the Association of
  Universities for Research in Astronomy, Inc., under cooperative agreement
  with the National Science Foundation.} task
\texttt{noao.imred.crutil.cosmicrays} to remove cosmic rays from the low
spectral resolution data and the L.A.Cosmic package to remove them from the
moderate spectral resolution data \citep{vanDokkum2001}. We determined the
dispersion solution using the \texttt{noao.twodspec.longslit.identify},
\texttt{reidentify}, \texttt{fitcoords}, and \texttt{transform} tasks and Ar
and Ne comparison lamp spectra for the low and moderate spectral resolution
observations, respectively. The heliocentric velocity correction was performed
using the \texttt{astutil.rvcorrect} task.

\begin{figure}[h]
\epsscale{1.2}\plotone{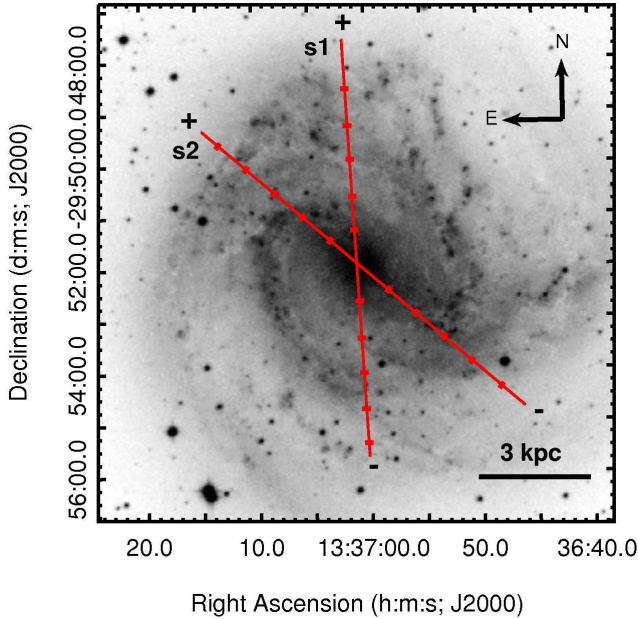}
\caption{The two longslits overlaid on a red image of M83 from the Digitized
  Sky Survey (Second Generation;
  https://archive.stsci.edu/cgi-bin/dss\_form). The tickmarks indicate
  galactocentric radii of $|R| = 1 - 5$ kpc, and the plus and minus signs show
  the sides of the galaxy with positive and negative $R$, respectively.}
\end{figure}

During each night of observations, we obtained a single separate sky exposure
that approximates the track followed by the telescope during the science
exposures, and scaled the former by a multiplicative factor to account for
variations in sky brightness. After sky subtraction, we combined the spectra
within, and then between, nights; to do so, we scaled the spectra by their
median values, stacked them by their median once again, and extracted them
using an aperture of 11 pixels. The aperture width was chosen to be large
enough to minimize the effect of curvature of the spatial axis with respect to
the pixel rows and to gain in signal-to-noise (S/N) without significantly
sacrificing the spatial resolution. We calculated the error bars from the rms
uncertainty in the continuum assuming that the S/N scales according to Poisson
statistics ($\sqrt{N}$). Vignetting of as much as 25$\%$ of the slit required
cropping of some frames before stacking, reducing the S/N at the ends of the
slits.

\subsection{Flux Calibration}\label{sec_phot_cal}

We perform the relative flux calibration using observations of the
spectrophotometric standard star Hiltner 600. Absolute flux calibration is not
possible with SALT alone due to the varying effective telescope area as a
function of time. Thus, to perform the flux calibration, we use \textit{Hubble
  Space Telescope (HST)} Wide Field Camera 3 imaging of M83 from the Hubble
Legacy Archive\footnote[3]{\url{https://hla.stsci.edu/}} (PI: Blair; Proposal
ID: 12513). The images are taken with the f657n narrow-band filter; at the
redshift of M83, the filter window includes the [NII]$\lambda\lambda$6548,
6583 and H$\alpha$ lines. The image mosaic covers $\sim 100\% \text{ and }
75\%$ of the spatial extents of slits 1 and 2, respectively.

To perform the flux calibration, we convolved the \textit{HST} mosaic with a
two-dimensional Gaussian using the IDL function \texttt{gauss smooth}. The
standard deviation of the Gaussian kernel was equal to the estimated seeing at
the SALT site, $\sigma = 1.5"$. For each section of the slit from which a
spectrum was extracted, we calculated the average flux density required to
produce the observed \textit{HST} counts using the PHOTFLAM
keyword. Accounting for the filter throughput, we then determined the average
flux density of the SALT spectra to yield a conversion factor between
instrumental and astrophysical units at H$\alpha$.

Due to saturation and scattered light in the nucleus, and vignetting at the
ends of the slits, we calculated a median conversion factor, $f$, for spectra
between $1 \text{ kpc} \le |R| \le 4 \text{ kpc}$\footnote[4]{Here and
  throughout this paper, $R$ refers to the true galactocentric radius, while
  $R'$ refers to the projected radius, where $R = R'
  \sqrt{\text{cos}^{2}(\phi) +
    \text{sin}^{2}(\phi)/\text{cos}^{2}(i)}$. Negative values of $R$
  correspond to the southwest side of the galaxy, and positive values of $R$
  to the northeast side.}. This yielded $f = 2.1 \pm 0.4 \times 10^{-16}
\text{ erg cm}^{-2} \text{ SALT ADU}^{-1}$ for slit 1 and $f = 2.3 \pm 0.4
\times 10^{-16} \text{ erg cm}^{-2} \text{ SALT ADU}^{-1}$ for slit 2,
suggesting that the flux calibration is accurate to $\sim 20\%$.

\subsection{Instrumental Scattered Light}\label{sec_inst_sl}

Within $\sim 40"$ ($|R| = 1$ kpc) of the center, an instrumental scattered
light feature appears as very broad emission ($\sigma = 0.5 - 1 \times 10^{3}$
\kms) under the H$\alpha$ and [NII]$\lambda\lambda$6548, 6583 emission
lines. This feature is most noticeable in the moderate spectral resolution
data at $|R| = 1$ kpc, where its intensity becomes comparable to the H$\alpha$
intensity. To verify that it is due to instrumental scattered light, we
obtained another RSS longslit observation using the same instrument setup, and
adjusted the pointing center (R.A. = 13 37 01.9, Decl. = -29 52 13, J2000) and
position angle (P.A. = 28$^{\circ}$) to avoid the brightest part of the
nucleus. The absence of the broad emission feature from this observation
suggests that it is due to instrumental scattered light from bright nuclear
star clusters.

We simultaneously remove the scattered light and the continuum by masking the
emission lines, smoothing with a Gaussian filter, and subtracting the
result. At low spectral resolution, we use a Gaussian window with $\sigma =
12.5$ \AA. At moderate spectral resolution, we choose $\sigma = 25$ \AA \ and
$\sigma = 10$ \AA; the latter is used where the scattered light intensity is
comparable to the H$\alpha$ intensity. These choices produce sufficiently
smooth continua while capturing the curvature of the scattered light emission
where necessary. We assume no additional error associated with the scattered
light and continuum subtraction.

\section{Detection of Multiple Emission-Line Components: A Markov Chain Monte Carlo Method}\label{sec_mcmc}

\begin{figure*}[h]
\epsscale{1.0}\plotone{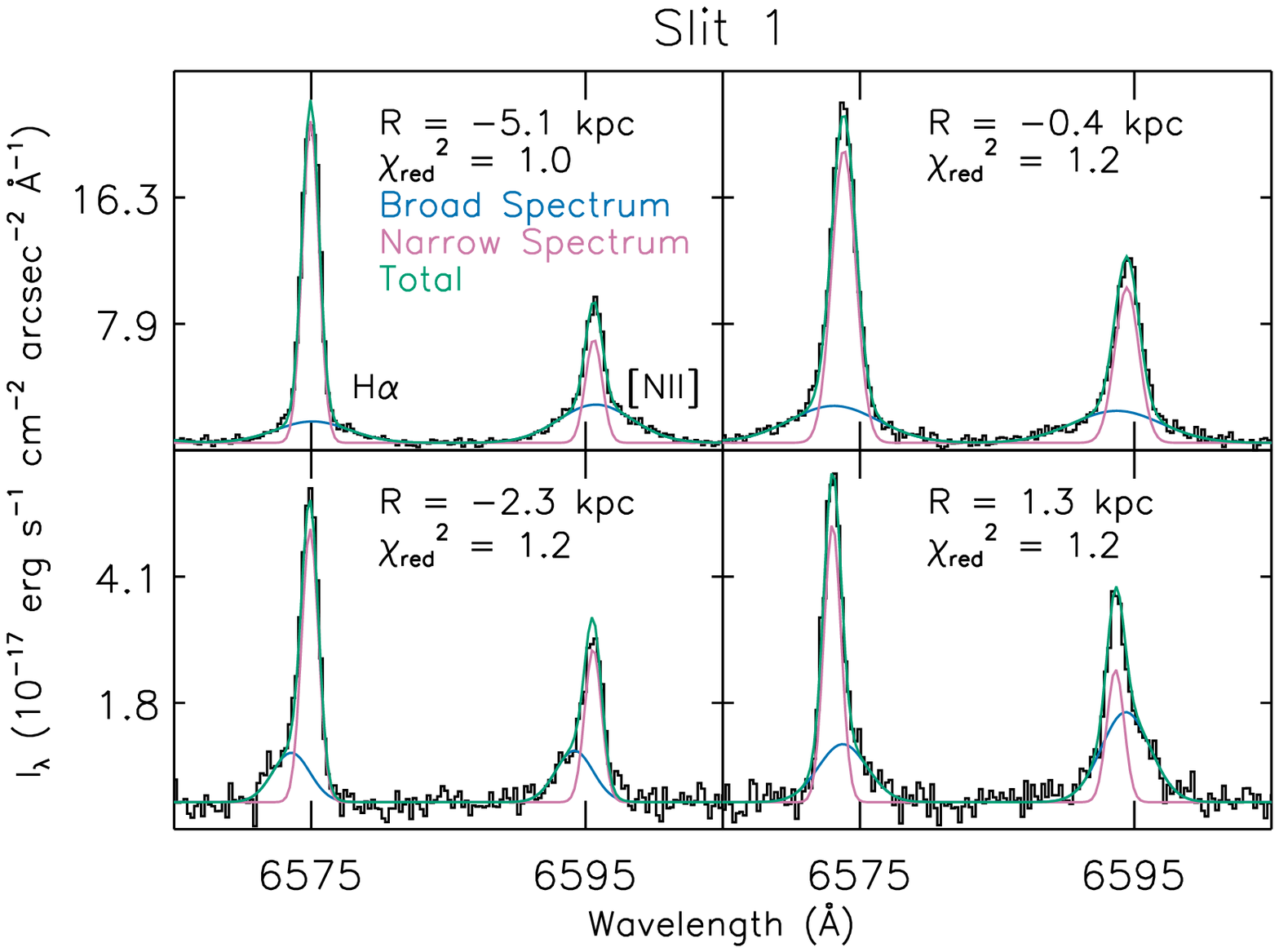}
\epsscale{1.0}\plotone{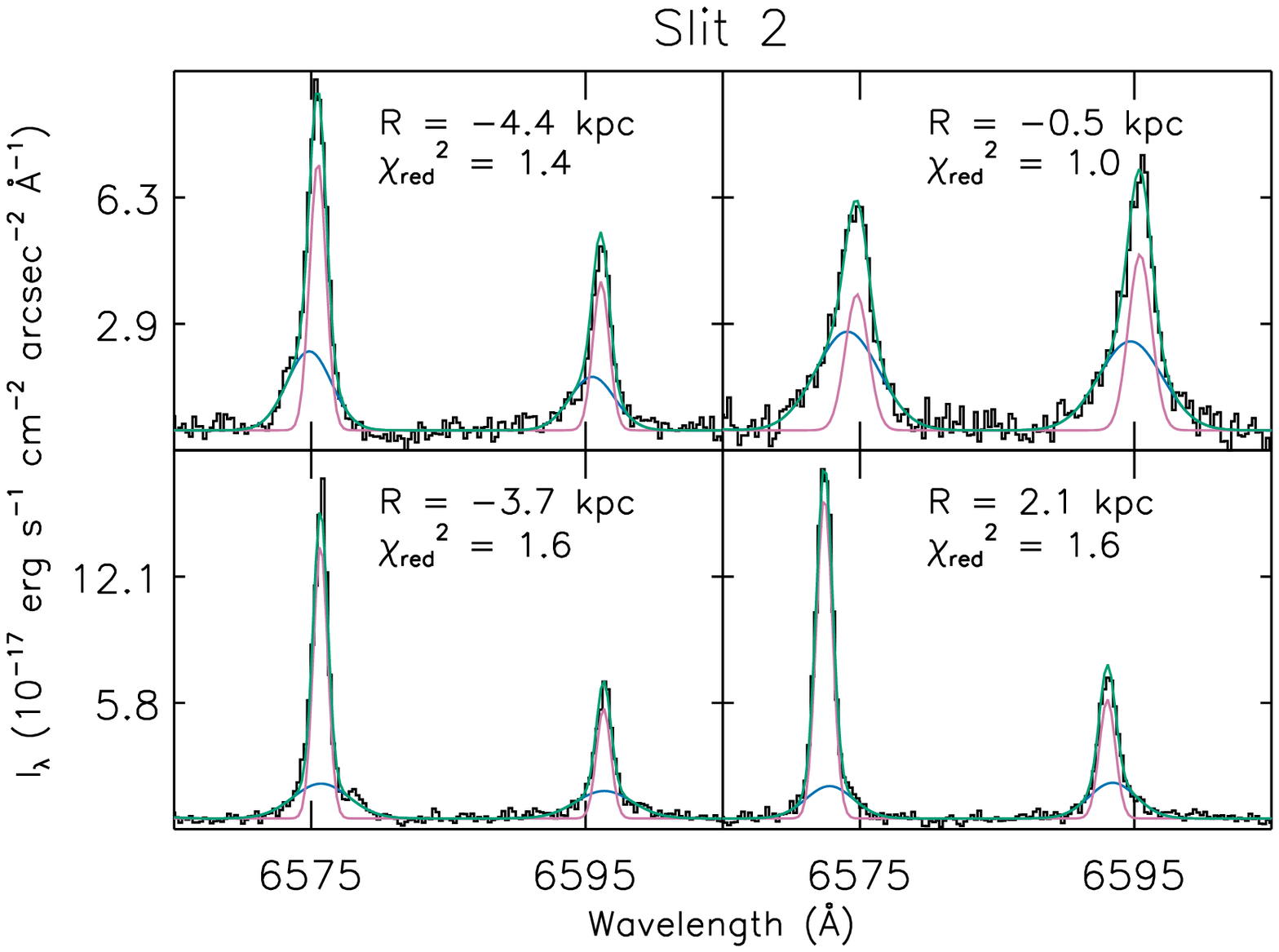}
\caption{Example spectra from slits 1 (top) and 2 (bottom) showing a Gaussian
  decomposition into a narrow (pink) and broad (blue) component. A Markov
  Chain Monte Carlo method was used to decompose the
  [NII]$\lambda\lambda$6548, 6583, H$\alpha$, and [SII]$\lambda\lambda$6717,
  6731 emission lines into multiple components; for clarity, only the
  H$\alpha$ and [NII]$\lambda$6583 lines are shown here. The spectra shown are
  chosen to illustrate the range of emission line intensities, morphologies,
  and ratios observed in each component.}
\end{figure*}

\begin{deluxetable*}{ccccccccc}[t]
\tabletypesize{\scriptsize}
\tablecolumns{9}
\tablewidth{0pt}
\tablecaption{Markov Chain Monte Carlo Parameters}
\tablehead{ 
\colhead{} &
\colhead{$I(H\alpha)_{b}/I(H\alpha)_{tot}$ \tablenotemark{a}} &
\colhead{[NII]$\lambda$6583/H$\alpha|_{b}$} &
\colhead{[SII]$\lambda$6717/H$\alpha|_{b}$} &
\colhead{[SII]$\lambda$6731/H$\alpha|_{b}$} &
\colhead{$v_{n}$} &
\colhead{$v_{b}$} &
\colhead{$\sigma_{n}$ \tablenotemark{b}} &
\colhead{$\sigma_{b}$}\\
\colhead{} &
\colhead{} &
\colhead{} &
\colhead{} &
\colhead{} &
\colhead{\kms} &
\colhead{\kms} &
\colhead{\kms} &
\colhead{\kms}
}
\startdata
Initial Value & 0.25 & 1.0 & 0.5 & 0.5 & 500 & 500 & 30 & 50 \\
Step Size & 0.1 & 0.1 & 0.1 & 0.1 & 10 & 10 & 10 & 10
\enddata
\tablenotetext{a}{The subscripts $n$ and $b$ refer to the narrow and broad
  components, respectively.}
\tablenotetext{b}{$\sigma$ refers to the standard deviation of the Gaussian,
  and not to the full width at half maximum (FWHM).}
\end{deluxetable*}

Our first goal is to identify multiple emission-line components.  Using the
moderate spectral resolution data, we model the [NII]$\lambda\lambda$6548,
6583, H$\alpha$, and [SII]$\lambda\lambda$6717, 6731 emission lines as a
superposition of two Gaussians, and ask whether these components are
consistent with arising from eDIG, planar DIG (pDIG), or HII regions. To do
so, we use the following criteria: 1) The [NII]$\lambda$6583/H$\alpha$ and
[SII]$\lambda$6717/H$\alpha$ emission-line ratios are higher in diffuse gas
than in HII regions \citep[e.g.,][]{Rand1998, Otte2002, Madsen2006}, 2) the
velocity dispersion in diffuse gas may be higher than in HII regions
(\citealt{Heald2006a, Heald2006b, Heald2007}; B16), and 3) eDIG may display a
rotational velocity lag with respect to the HII regions in the disk
\citep[e.g.,][]{Fraternali2004, Heald2006a, Heald2006b, Heald2007,
  Bizyaev2017}. Since adding additional Gaussians will almost always improve
the quality of the fit, these considerations are crucial to tie the Gaussian
decomposition to the underlying physical processes. Note that here and
throughout the rest of this paper, HII region emission refers not only to that
from individual Str{\"o}mgren spheres, but also to that from the planar,
dense, ionized gas found locally in star-forming regions.

We use a Markov Chain Monte Carlo (MCMC) method with a Metropolis-Hastings
algorithm to model the emission-line profiles as the sum of a narrow and a
broad component \citep[e.g.,][]{Ivezic2014}. We probe an 8-dimensional
parameter space defined by the broad H$\alpha$ intensity,
[NII]$\lambda$6583/H$\alpha$, [SII]$\lambda$6717/H$\alpha$, and
[SII]$\lambda$6731/H$\alpha$, as well as the broad and narrow velocities and
velocity dispersions. To prevent degeneracies in parameter space, we require
that the velocity dispersion of the narrow component not exceed that of the
broad. We assume that the velocity dispersions do not depend on atomic
species; as we will see in \S4.2, this is a reasonable assumption since the
turbulent contribution dominates the thermal contribution. We also assume that
[NII]$\lambda$6548/[NII]$\lambda$6583 = 0.3 for both the broad and narrow
component, but do not make an assumption about the value of
[SII]$\lambda$6717/[SII]$\lambda$6731. We require that the sum of the narrow
and broad intensities equal the observed, integrated intensity. The analysis
is only performed if at least three of the five emission lines are detected at
the 5$\sigma$ level.

For a given spectrum, the MCMC method is implemented as follows. First, we
chose a location in parameter space, construct a model, and quantify the
quality of the fit using the $\chi^{2}$ statistic:
\begin{equation}
\chi^{2} = \sum_{n} \frac{(f_{\lambda,obs} - f_{\lambda,mod})^{2}}{\sigma_{obs}^{2}},
\end{equation}
where $f_{\lambda,obs}$ and $f_{\lambda,mod}$ are the observed and modeled
flux densities, $\sigma_{obs}$ is the observed uncertainty, and $n$ is the
number of wavelength bins. We calculate the value of $\chi^{2}$ within 8 \AA \
of the center of the emission lines of interest.

Next, we select a parameter from a uniform distribution. We then select a
distance and direction to move in that parameter from a Gaussian distribution
with a standard deviation equal to a given step size. The decision to accept
or reject this new model is based on an acceptance probability given by:
\begin{equation}
p = \text{e}^{-(\chi_{new}^{2} - \chi_{old}^{2})/2},
\end{equation}
where $\chi_{new}^{2}$ and $\chi_{old}^{2}$ give the quality of the fit of the
new and old models, respectively. If $p > 1$, the new model is accepted as a
better fit. If $p < 1$, the value of $p$ is compared to a number $n$ from a
uniform distribution where $0 \le n \le 1$. If $p > n$, we accept the new
model. Likewise, if $p < n$, we reject the new model in favor of the old. This
ensures that a new model is always accepted when it provides a better fit, and
is sometimes accepted when it doesn't to ensure sufficient sampling of
parameter space.

This process is repeated for $N = 10^{5}$ links in the MCMC chain; our choice
of $N$ is discussed in Appendix A, where we demonstrate the convergence of the
algorithm. The first $10^{4}$ links are rejected as the ``burn-in''. The
remaining links are used to construct distributions of the accepted values of
each parameter. The median values of the distributions are taken as the
parameter values, and the median absolute deviations are taken as the
parameter uncertainties. In Table 2, we show the starting values and step
sizes of each parameter. The results are not sensitive to either the start
values or the step sizes; however, we select the step sizes to achieve an
acceptance fraction of $ < 50\%$ to ensure effective sampling of the parameter
space.

As a check on our parameter uncertainties, we randomly perturb each flux bin
in every spectrum by adding or subtracting the one-sigma error bar. We then
run the same MCMC algorithm on the perturbed data, and compare the best-fit
parameters from the perturbed and unperturbed data. The uncertainty estimate
implied from this comparison exceeds that implied by the original MCMC
analysis for all parameters. For most parameters, the effect is small ($10 -
30\%$); however, in some cases, it is as high as a factor of two. Thus, when
considering the error estimates in this paper, it should be noted that they
may be underestimated by factors in this range.

After the MCMC algorithm constructs a two-component model for each spectrum,
we implement an additional criterion to identify the true two-component
spectra. In cases where the emission lines are well-represented by a single
Gaussian, the second Gaussian often fits to noise, structure in the continuum,
or curvature caused by the H$\alpha$ stellar absorption feature. To identify
true two-component spectra, we require that each component comprises at least
15$\%$ of the integrated intensity of at least 3 of the emission lines. Those
without two-component spectra are fit with single Gaussians, yielding both
multi-component (broad, narrow) and single-component fits. Of 388 total
spectra, 350 spectra have 5$\sigma$ detections, and 191 and 159 spectra have
multi- and single-component fits, respectively. The multi-component spectra
are distributed across the full range of galactocentric radii considered.

In Figure 2, we show a comparison of the best-fit two-component modeled and
observed spectra. By eye, the quality of the fits are very good, and the
reduced $\chi^{2}$ values of the two-component fits have a median value of
$\chi_{red}^{2} = 1.4$. Due to the degeneracy of Gaussian decomposition, this
approach cannot produce a unique decomposition for a given spectrum. However,
taken in aggregate, the results reveal the physical conditions and kinematics
in the narrow- and broad-line emitting regions. We discuss the properties of
the narrow and broad components, and their relationship to eDIG, pDIG, and HII
region emission, in the following section.

\section{Observational Results}\label{sec_results}

\subsection{Identification of DIG Emission}\label{sec_prop}

\begin{figure}[h]
\epsscale{1.2}\plotone{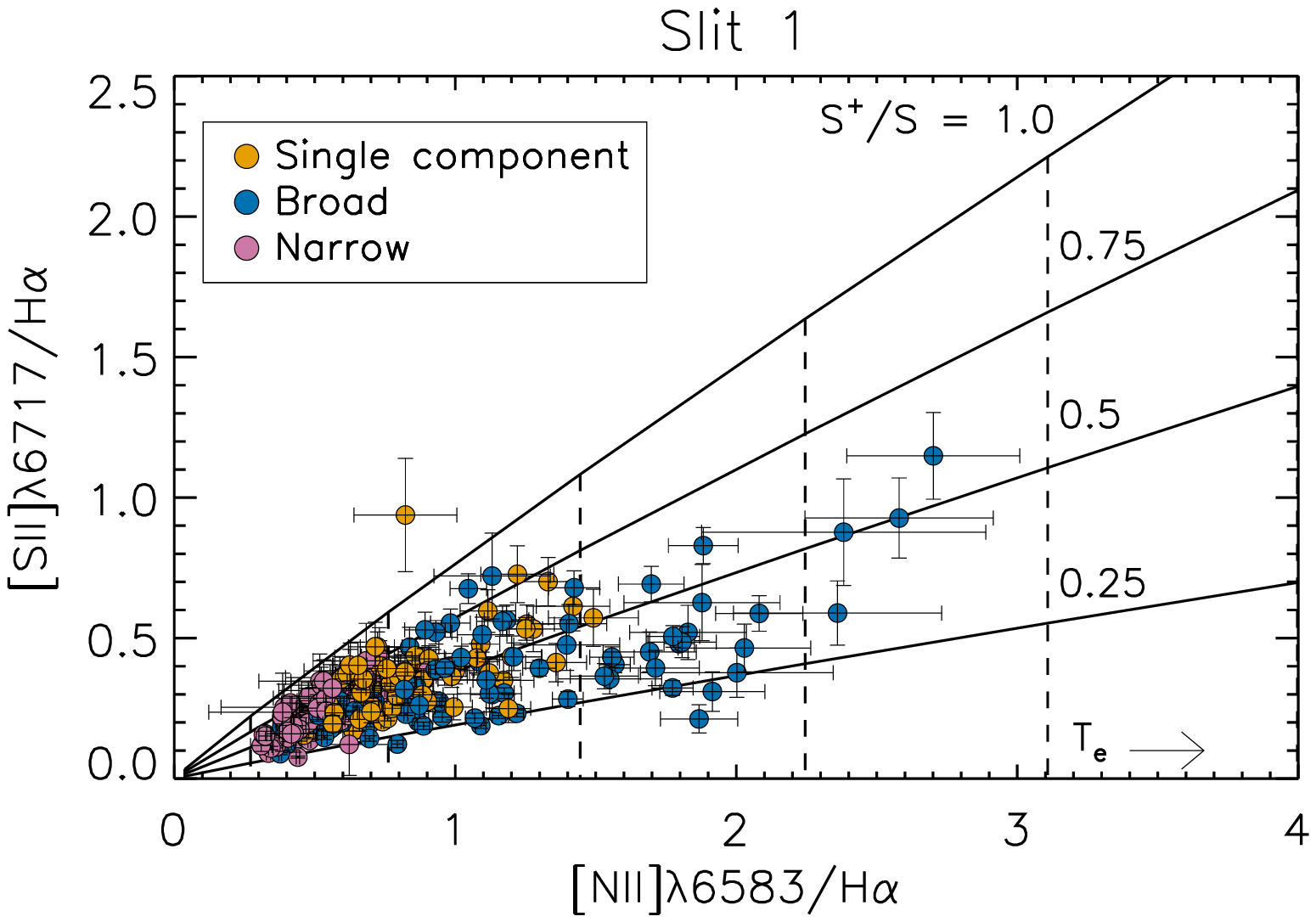}
\epsscale{1.2}\plotone{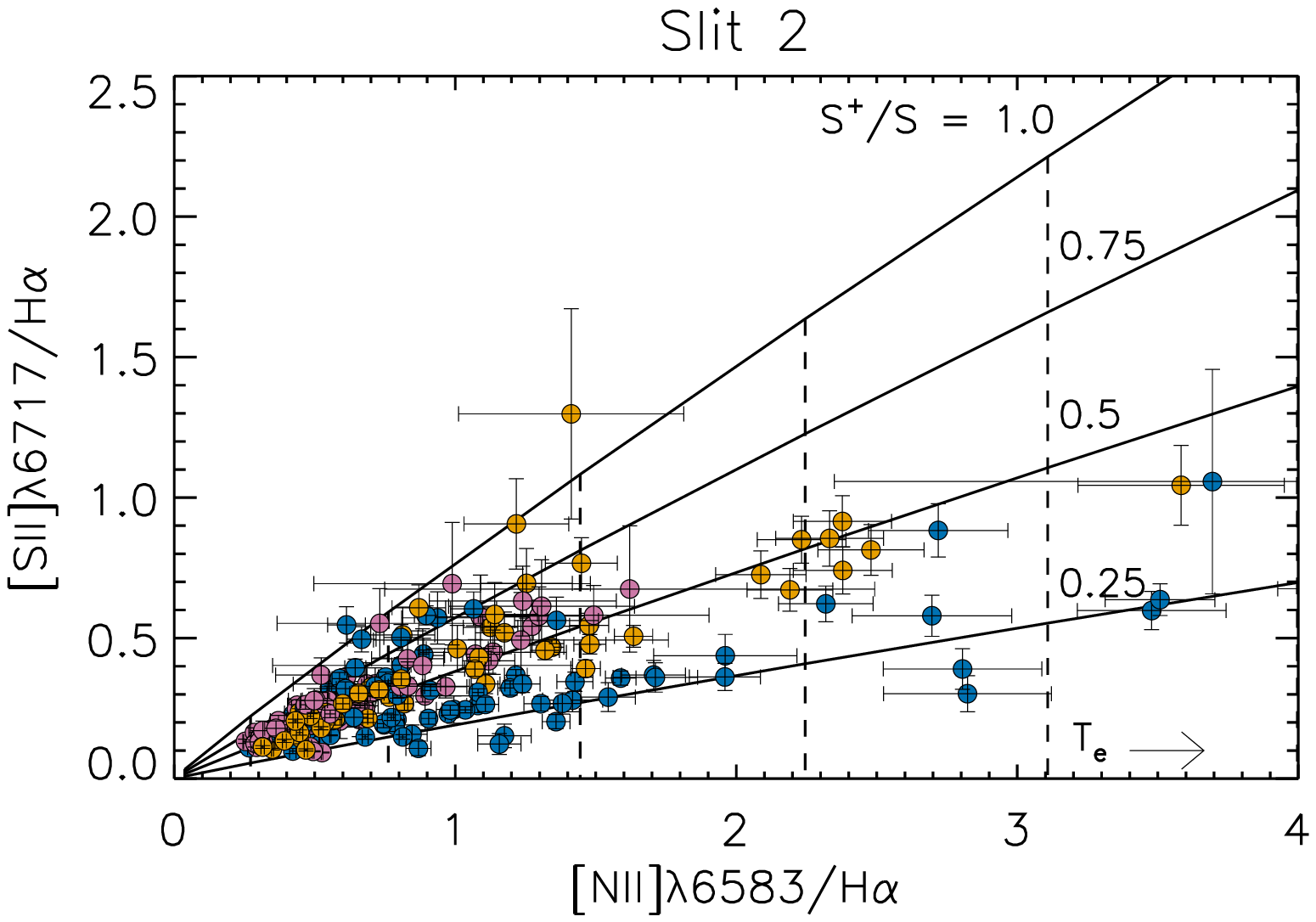}
\caption{The narrow (pink) and broad (blue) emission-line components from
  slits 1 (top) and 2 (bottom) lie in regions of the
  [NII]$\lambda$6583/H$\alpha$, [SII]$\lambda$6717/H$\alpha$ plane consistent
  with HII region and DIG emission, respectively. The solid lines correspond
  to a constant ionization fraction of $S^{+}/S = 0.25, 0.5, 0.75, \text{and }
  1.0$ from bottom to top, and the dashed lines indicate a constant electron
  temperature of $T_{e} = 0.6, 0.8, 1.0, 1.2, \text{and } 1.4 \times 10^{4}$ K
  from left to right. The broad components lie largely at high ratios of
  forbidden line emission to recombination line emission suggestive of high
  temperatures ($0.8 \times 10^{4}$ K $< T_{e} < 1.4 \times 10^{4}$ K) and
  ionization states ($S^{+}/S < 0.75$). The single-component spectra (yellow)
  tend to arise from intermediate physical conditions; however, in slit 2,
  single-component fits with high [NII]$\lambda$6583/H$\alpha$ are found
  within $|R| < 2$ kpc where there is evidence for shock ionization near
  star-forming regions \citep{Calzetti2004, Hong2011}. In slit 2, seven data
  points lie at [NII]$\lambda$6583/H$\alpha > 4$.}
\end{figure}

In Figure 3, we compare the [NII]$\lambda$6583/H$\alpha$ and
[SII]$\lambda$6717/H$\alpha$ emission-line ratios for the narrow, broad, and
single-component spectra. We indicate the expected values of these
emission-line ratios based on abundances, electron temperature, and ionization
fraction as follows. The emission-line ratios can be expressed as:
\begin{equation}
\begin{split}
\frac{I([NII]\lambda 6583)}{I(H\alpha)} = 1.63 \times 10^{5}
\bigg(\frac{H^{+}}{H} \bigg)^{-1} \bigg(\frac{N}{H}\bigg)
\bigg(\frac{N^{+}}{N}\bigg) \times \\ T_{4}^{0.426} \text{e}^{-2.18/T_{4}}
\end{split}
\end{equation}
and
\begin{equation}
\begin{split}
\frac{I([SII]\lambda 6717)}{I(H\alpha)} = 7.67 \times 10^{5}
\bigg(\frac{H^{+}}{H} \bigg)^{-1} \bigg(\frac{S}{H}\bigg)
\bigg(\frac{S^{+}}{S}\bigg) \times \\ T_{4}^{0.307} \text{e}^{-2.14/T_{4}},
\end{split}
\end{equation}
where $T_{4}$ is the electron temperature in units of $10^{4}$ K
\citep{Haffner1999, Osterbrock2006}. We use $N/H = 9.8 \times 10^{-5}$ and
$S/H = 1.2 \times 10^{-5}$ as median values of direct abundances determined
from auroral lines detected in five HII regions by \citet[][see their Table
12]{Bresolin2005}. We assume that the H and N are $100\%$ and $80\%$ ionized,
respectively; in the Milky Way, $H^{+}/H > 0.9$ for $T_{4} > 0.8$
\citep{Reynolds1998}, and $N^{+}/N \sim 0.8$ under a range of DIG conditions
\citep{Sembach2000}. Though $N^{+}/N$ is fairly constant in the DIG, $S^{+}/S$
may vary under these conditions, due to the different second ionization
potentials of these species (23.3 eV for $S^{+} \rightarrow S^{2+}$, 29.6 eV
for $N^{+} \rightarrow N^{2+}$).

In Figure 3, we allow both $T_{4}$ and $S^{+}/S$ to vary, indicating dashed
lines of constant $T_{4}$ and solid lines of constant $S^{+}/S$. This analysis
neglects, among other things, a radial abundance gradient and variations in
abundances, ionization fractions, and electron temperature along the line of
sight. However, it gives us a qualitative sense of the physical conditions
from which the narrow and broad emission arise.

\begin{figure*}[h]
\epsscale{1.2}\plotone{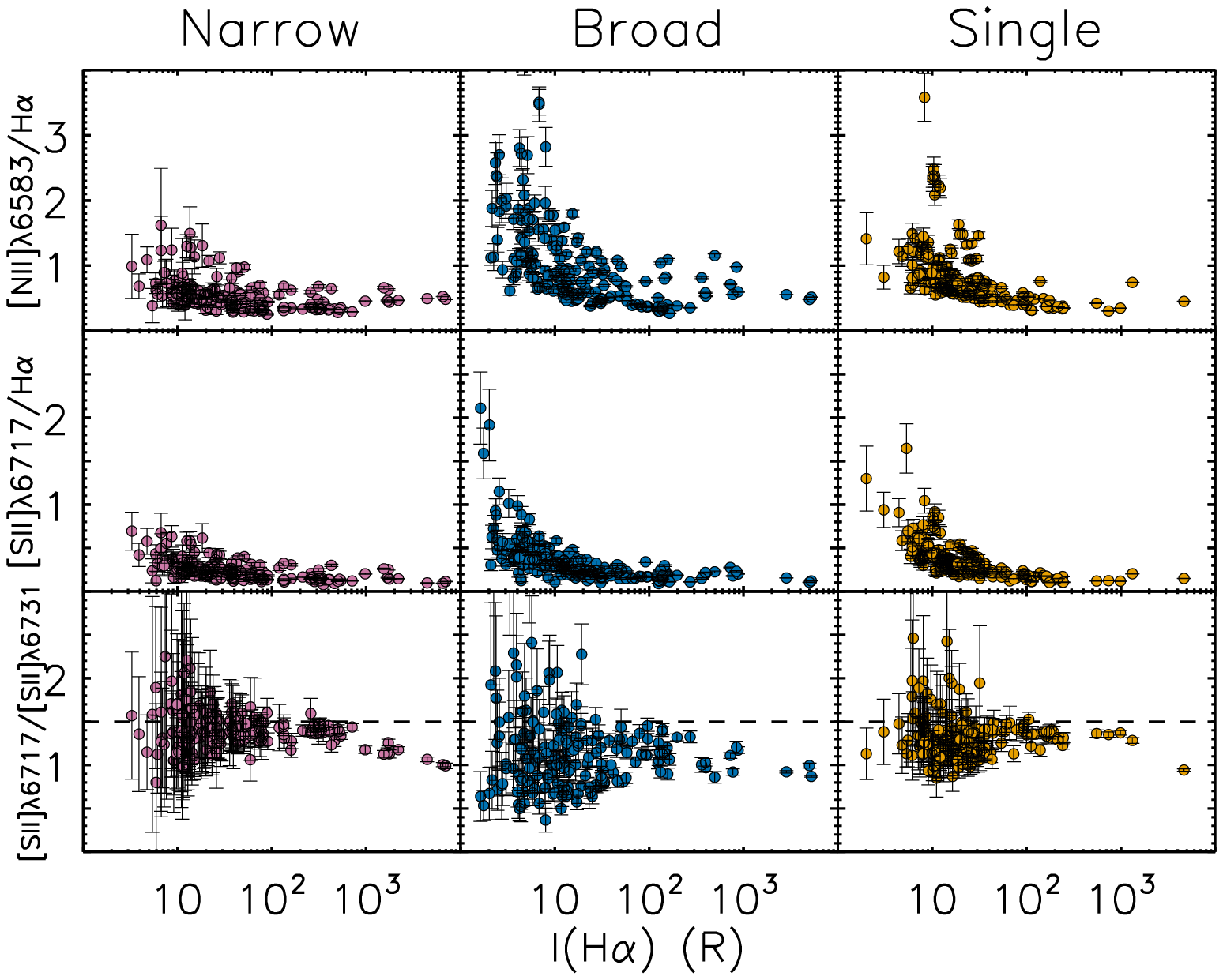}
\caption{[NII]$\lambda$6583/H$\alpha$ (top row),
    [SII]$\lambda$6717/H$\alpha$ (middle), and
    [SII]$\lambda$6717/[SII]$\lambda$6731 (bottom) as functions of
    $I(H\alpha)$ for the narrow (left), broad (center), and single-component
    (right) spectra. The surface brightness is corrected for the inclination
    of the galaxy. \textit{Top and middle rows:} The broad emission tends to
    fall at fainter $I(H\alpha)$ and higher [NII]$\lambda$6583/H$\alpha$ and
    [SII]$\lambda$6717/H$\alpha$ than the narrow emission, consistent with
    arising from a more diffuse medium subject to supplemental heating (i.e.,
    a heating mechanism proportional to $n_{e}^{\alpha}$, where $\alpha <
    2$). \textit{Bottom row:} The narrow and single components are generally
    consistent with the low-density limit within the errors
    ([SII]$\lambda$6717/[SII]$\lambda$6731 $= 1.5$, indicated by the dashed
    line). However, the broad component shows a much larger scatter in the
    line ratio; while measurements at $I(H\alpha) \ge 10^{2} \text{ R}$ may be
    indicative of dense shells in the starburst nucleus, those at $I(H\alpha)
    \le 30 \text{ R}$ are unreliable due to low S/N and possible instrumental
    effects.}
\end{figure*}

In slit 1, the narrow emission lies at the lowest values of the emission-line
ratios around [NII]$\lambda 6583$/H$\alpha \sim 0.5$ and [SII]$\lambda
6717$/H$\alpha \sim 0.3$ between $0.6 \le T_{4} \le 0.9$. In contrast, the
broad emission is scattered between $0.5 \le$ [NII]$\lambda 6583/$H$\alpha \le
2.7$, $0.1 \le$ [SII]$\lambda 6717/$H$\alpha \le 1.2$, and $0.7 \le T_{4} \le
1.3$. Both components span a range in ionization fraction ($0.25 \le S^{+}/S
\le 1.0$). Thus, the narrow and broad emission are consistent with arising
from HII regions and the warmer DIG in a metal rich galaxy; a similar behavior
is seen for emission from HII regions and WIM in the Milky Way
\citep[e.g.,][]{Haffner1999}. The single-component spectra show intermediate
line ratios between $0.7 \le T_{4} \le 1.0$, suggesting that these spectra
originate from a range of physical conditions. Those with lower and higher
emission-line ratios are likely dominated by HII region and DIG emission,
respectively.

In slit 2, the spectra show a similar behavior, although considerably more
single-component spectra lie at [NII]$\lambda 6583$/H$\alpha > 1.5$, and more
broad spectra are found at [NII]$\lambda 6583$/H$\alpha > 2$. These spectra
are largely found at $|R| < 2$ kpc, where LINER-like emission-line ratios have
been previously observed. This emission is likely due to shock ionization from
stellar winds and supernovae near star-forming regions \citep{Calzetti2004,
  Hong2011}. As discussed in \S2.2, this phenomenon is spatially coincident
with instrumental scattered light, and it is possible that residuals from
scattered light subtraction interfere with our ability to identify multiple
components.

As shown in the top and middle rows of Figure 4, [NII]$\lambda 6583$/H$\alpha$
and [SII]$\lambda 6717/$H$\alpha$ increase as $I(H\alpha)$ decreases. Thus,
the fainter, broad emission, with H$\alpha$ intensities largely between $1
\text{ R} < I(H\alpha) < 10^{2} \text{ R}$, has higher ratios of forbidden
line to recombination line emission than the brighter, narrow emission with
$10 \text{ R} < I(H\alpha) < 10^{3} \text{ R}$. This trend, observed in the
Milky Way and other nearby, edge-on disk galaxies \citep[e.g.,][]{Rand1998,
  Haffner1999}, has been attributed to a supplemental heating mechanism at low
electron density, $n_{e}$, proportional to $n_{e}^{\alpha}$ for $\alpha < 2$
(i.e., proportional to a lower power of $n_{e}$ than photoionization heating,
$n_{e}^{2}$). Note that bright, broad emission with low emission-line ratios
is observed in the nucleus of M83, likely due to dense, planar gas surrounding
star-forming regions.

In the bottom row of Figure 4, we compare the
[SII]$\lambda$6717/[SII]$\lambda$6731 emission-line ratios of the narrow,
broad, and single components as a function of $I(H\alpha)$. As expected, the
narrow and single components are largely at the low-density limit of
[SII]$\lambda$6717/[SII]$\lambda$6731 $= 1.5$. For $I(H\alpha) > 10^{2}$, the
broad component has [SII]$\lambda$6717/[SII]$\lambda$6731 $\sim 1$, consistent
with electron densities of $n_{e} = 10^{2} - 10^{3} \text{ cm}^{-3}$
\citep{Osterbrock2006}. This may be indicative of dense shells and filaments
associated with the nuclear starburst. At $I(H\alpha) \le 30 \text{ R}$, there
is significant scatter in [SII]$\lambda$6717/[SII]$\lambda$6731, and we do not
trust the measured line ratios in this regime. The S/N is insufficient to
robustly measure the line ratios at these intensities, and the results may be
influenced by instrumental effects. A high S/N measurement of
[SII]$\lambda$6717/[SII]$\lambda$6731 at faint $I(H\alpha)$ is of interest to
determine if any of the broad emission originates from dense gas.

We do not correct for the H$\alpha$ stellar absorption line. The relative
impact of absorption on the broad and narrow components is unclear. It is
possible that absorption impacts the intensity of the narrow component more
than the broad, as the former is more likely to be aligned with the stellar
population in velocity space. Nevertheless, we estimate the impact of
absorption if it exclusively impacts the broad and narrow components,
respectively. Assuming an H$\alpha$ absorption line with an equivalent width
of $EW = 2$ \AA, the H$\alpha$ absorption is generally comparable to the broad
H$\alpha$ intensity, and ranges from comparable to smaller by several orders
of magnitude for the narrow H$\alpha$ intensity. If the intensities are
corrected for this absorption, the maximum observed line ratios are
[NII]$\lambda$6583/H$\alpha \sim 1.5$ (0.5) and [SII]$\lambda$6717/H$\alpha
\sim 0.4$ (0.3) for the broad (narrow) components.

In general, the broad component is consistent with arising from diffuse gas,
while the narrow component is consistent with originating in HII regions. This
is supported by the former components' high [NII]$\lambda$6583/H$\alpha$ and
[SII]$\lambda$6717/H$\alpha$ line ratios at faint $I(H\alpha)$. In the next
section, we assess the kinematics of the narrow and broad components, and
consider evidence for an eDIG layer.

\subsection{Kinematics and Identification of eDIG Emission}\label{sec_kin}

In Figure 5, the line-of-sight velocity dispersion, $\sigma$, is shown as a
function of $I(H\alpha)$. Here and throughout the rest of this paper, $\sigma$
refers to the standard deviation of the Gaussian fit. The velocity dispersion
is corrected for instrumental resolution ($\sigma^{2} = \sigma_{obs}^{2} -
\sigma_{res}^{2}$). The narrow emission generally has $10 \text{\kms} < \sigma
< 30$ \kms, with a median value of $\sigma = 20$ \kms, consistent with HII
region line widths of a few tens of \kms. Widths as large as $\sigma = 50$
\kms are observed in the brightest narrow components from the turbulent,
star-forming nucleus.

The broad component has a remarkable median velocity dispersion of $\sigma =
96$ \kms, with a significant spread around this value of $40 \text{\kms} <
\sigma < 150$ \kms. The moderate and large velocity dispersions of the narrow
and broad components are suggestive of thin (planar) and thick (extraplanar)
gaseous disks. With a median line width of $\sigma = 26$ \kms, the
single-component spectra have widths much more comparable to the narrow
component than to the broad.

The heliocentric, line-of-sight velocities of the narrow, broad, and single
components are shown as a function of galactocentric radius in Figure 6. The
narrow and single components are dominated by the rotational velocity of the
disk, but the broad component tends toward systemic velocity. The median
difference in radial velocity between the narrow and the broad components
implies a rotational velocity lag of $\Delta v = -24$ \kms in projection, or
$\Delta v = -70$ \kms corrected for inclination.

This is qualitatively consistent with the rotational velocity lags that are
characteristic of multi-phase, gaseous halos \citep[e.g.,][]{Fraternali2002,
  Fraternali2004, Heald2006a, Heald2006b, Heald2007, Oosterloo2007,
  Bizyaev2017}, but is quantitatively in excess of the $\Delta v = -10 - -40$
\kms kpc$^{-1}$ commonly observed in nearby, edge-on eDIG layers
\citep{Heald2006a, Heald2006b, Heald2007, Bizyaev2017}. These lags are often
interpreted as evidence of a galactic fountain; as gas clouds rise out of the
disk, they experience a weaker gravitational field, move out in radius, and
slow down to conserve angular momentum \citep[e.g.,][]{Collins2002}.

\begin{figure}[h]
\epsscale{1.2}\plotone{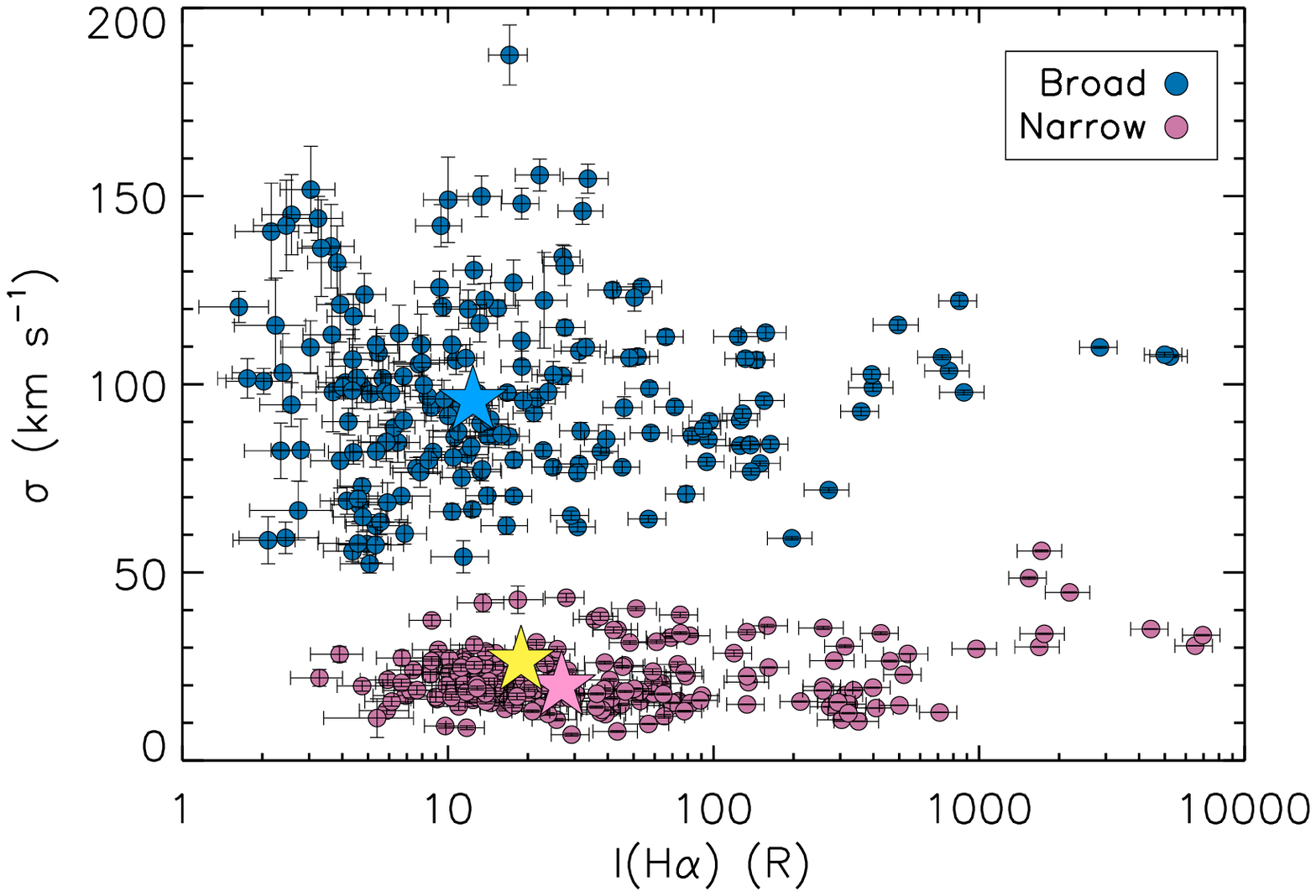}
\caption{The line widths of the broad components (blue) greatly exceed those
  of the narrow components (pink), suggesting that the former originates from
  a thicker gaseous disk. The median values of $\sigma$ and $I(H\alpha)$ for
  the broad ($\sigma = 96$ \kms), narrow ($\sigma = 20$ \kms), and single
  ($\sigma = 26$ \kms) components are shown with the blue, pink, and yellow
  stars, respectively. The brightest, broadest narrow components at
  $I(H\alpha) \ge 10^{3}$ R originate in the nucleus. The individual
  single-component spectra are not shown for visual clarity. The line widths
  are corrected for the instrumental resolution, and the surface brightness is
  corrected for the inclination of the galaxy.}
\end{figure}

There is also evidence of local bulk flows in the broad component. In slit 1,
outflows are suggested by the blueshifted gas near the nucleus ($-1 \text{
  kpc} < R < 0 \text{ kpc}$) and on the northeast side of the galaxy ($3
\text{ kpc} < R < 4 \text{ kpc}$). The most remarkable local feature in slit 2
is at $R = -4$ kpc, where coherent, redshifted velocities arise where the slit
crosses a star-forming spiral arm. These features may be due to expanding or
collapsing shells or other local bulk motions characteristic of a galactic
fountain flow.

One may ask whether the velocity profile of the broad component can be
explained by a series of local bulk flows alone. However, this requires a
preferential blueshifting and redshifting of the gas on the receding and
approaching sides of the galaxy, respectively. Lacking a physical basis for
this bias, the velocity profile is likely due to a lagging halo punctuated by
local bulk flows. In general, the large velocity dispersion, rotational
velocity lag, and local inflow and outflow are consistent with the broad
emission arising from an eDIG layer.

\begin{figure}[h]
\epsscale{1.2}\plotone{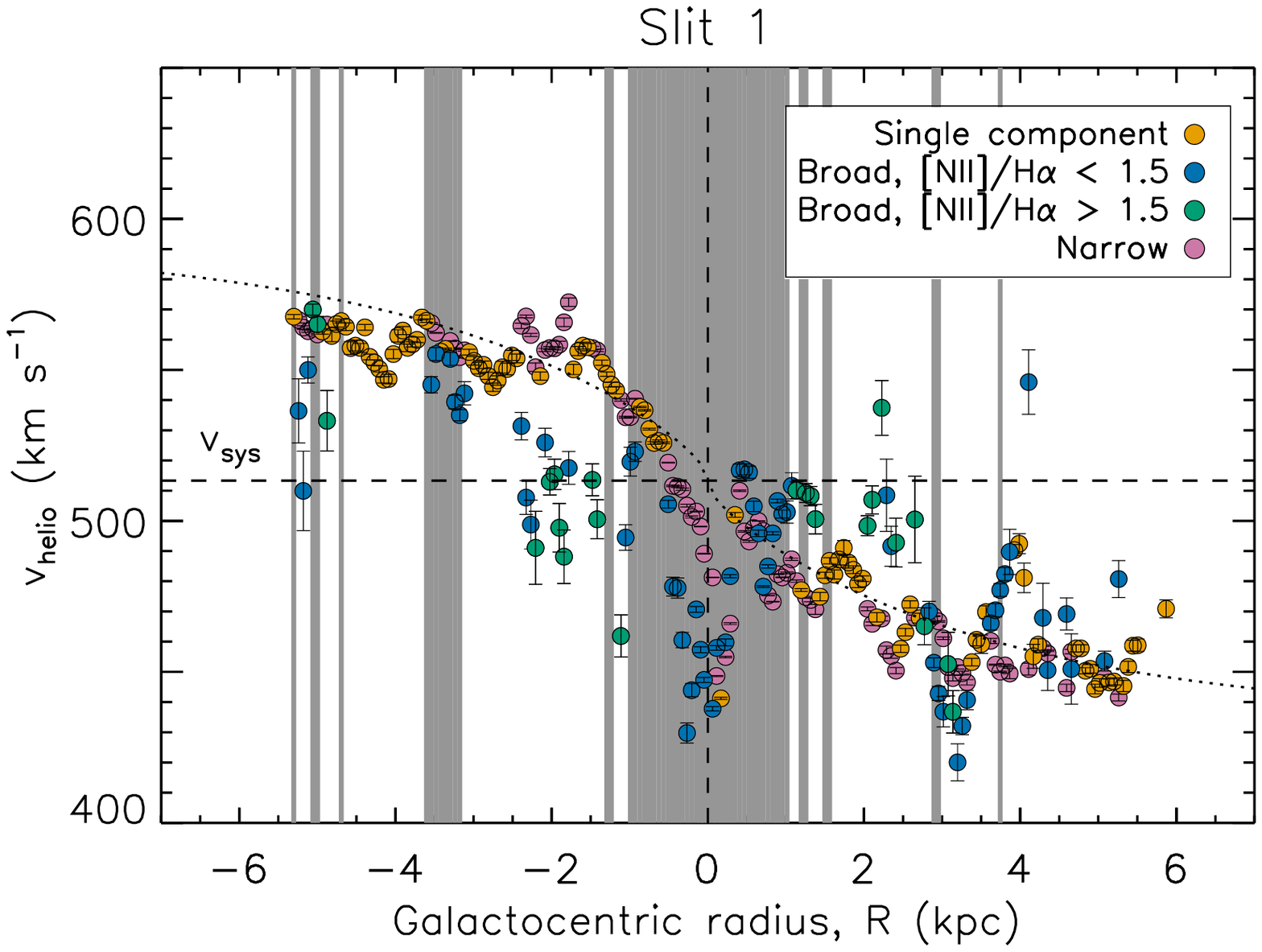}
\epsscale{1.2}\plotone{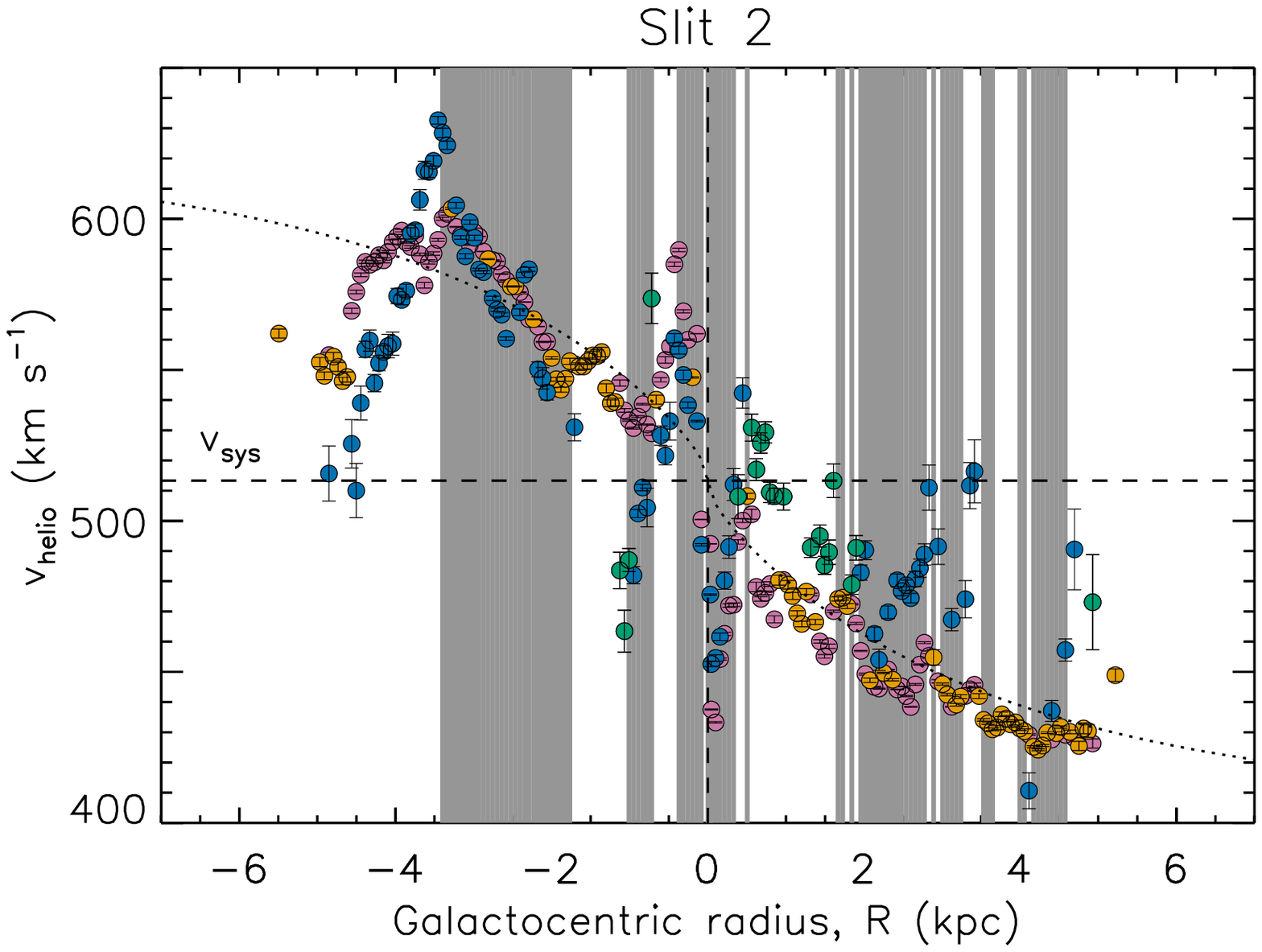}
\caption{For slits 1 (top) and 2 (bottom), the heliocentric radial velocities
  are shown for the narrow (pink), broad (blue and green), and
  single-component (yellow) spectra as functions of galactocentric
  radius. Broad emission with [NII]$\lambda 6583$/H$\alpha < 1.5$ and with
  [NII]$\lambda 6583$/H$\alpha > 1.5$ are shown in blue and green,
  respectively. Our best-fit rotation curve for M83 is shown by the dotted
  lines (see Appendix B), and the systemic velocity, $v_{sys} = 513$ \kms, is
  indicated by the dashed lines. The shaded radii indicate regions of
  star-formation activity (see \S4.3). In general, the broad component tends
  toward systemic velocity, suggesting an extraplanar gas layer with
  decreasing rotational velocity as a function of height above the disk. There
  is also evidence of local bulk flows; for example, in slit 1, a bulk
  blueshifting of the gas is seen near the nucleus ($-1 \text{ kpc} < R < 0
  \text{ kpc}$) and near $R = 3$ kpc.}
\end{figure}

\subsection{Proximity of eDIG Detection to Star-Formation
  Activity}\label{sec_mass}

Here, we evaluate the proximity of eDIG detection to star-formation activity
in the disk. In doing so, we ask where the broad component truly arises from
eDIG emission, and where it is due to pDIG emission associated with
star-forming spiral arms.
  
We consider an observation to be from a star-forming region if the narrow
$I(H\alpha)$ is at least three times higher than the minimum observed
$I(H\alpha) \sim 10$ R. In Figure 7, we shade the galactocentric radii that
meet this criterion. In the top panels, we show the broad $I(H\alpha)$ as a
function of $R$. It is clear that the faintest emission is detected away from
star-forming regions. $I(H\alpha)$ increases by several orders of magnitude in
the starburst nucleus, and by factors that range from a few to an order of
magnitude near star formation at larger $R$. This is a consequence of the
$n_{e}^{2}$ dependence of the H$\alpha$ intensity; if a bright, planar
component of the DIG exists along the line of sight, it will dominate the
broad emission-line profile and compromise our ability to detect a fainter,
extraplanar component along the same line of the sight.

In the bottom panels, we show the broad [NII]$\lambda 6583$/H$\alpha$ as a
function of $R$. The highest values of [NII]$\lambda 6583$/H$\alpha$ that are
indicative of the most diffuse gas are found between areas of star-formation
activity. This again suggests that the broad component is dominated by pDIG
emission near areas of star formation, and by eDIG emission elsewhere.

To confirm an eDIG detection, we must also consider the gas kinematics. In
Figure 6, we shade the star-forming regions on the position-velocity diagrams,
and we distinguish between broad emission in two emission-line ratio
regimes. The most diffuse, broad emission with [NII]$\lambda 6583$/H$\alpha >
1.5$ is shown in green, and the rest of the broad emission is shown in
blue. In general, the most diffuse emission tends toward systemic velocity,
consistent with a warm, ionized component of a lagging halo. The less diffuse
emission shows a range of kinematics; in some places, it is consistent with
the velocity of the disk, while in others it tends toward systemic velocity or
appears to be locally inflowing or outflowing.

In summary, the broad component arises from diffuse gas with a range of
densities and proximities to sources of ionizing radiation. Away from
star-forming regions, we detect broad emission with the clearest signature of
extraplanar, diffuse gas: faint $I(H\alpha)$, high values of [NII]$\lambda
6583$/H$\alpha$, and a rotational velocity lag with respect to the disk. Close
to star-formation activity, the broad emission is suggestive of planar,
diffuse gas or the base of the eDIG layer: brighter $I(H\alpha)$, lower values
of [NII]$\lambda 6583$/H$\alpha$, and more complex kinematics that include
rotation with the disk and local bulk flows. This result does not preclude the
possibility of an eDIG layer found above both star-forming and quiescent
regions. However, we cannot detect extraplanar emission along lines of sight
dominated by bright, broad, planar emission with the methods used here.

Note that the line width of the broad component does not show a clear trend
with $I(H\alpha)$, emission-line ratios, or line-of-sight velocity, so we take
the median line width of the broad emission as the velocity dispersion of the
eDIG layer for the remainder of this paper.

\begin{figure*}[h]
\epsscale{1.0}\plotone{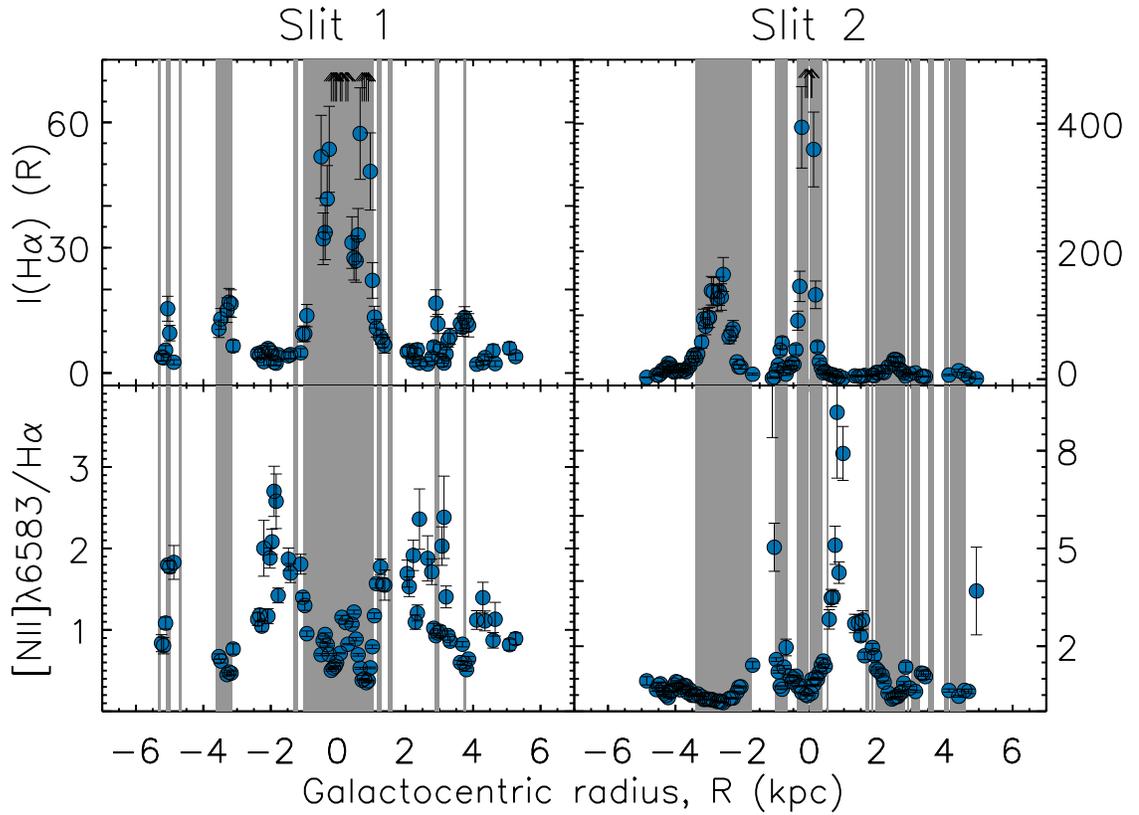}
\caption{The H$\alpha$ intensities and [NII]$\lambda 6583$/H$\alpha$
  emission-line ratios of the broad component as functions of galactocentric
  radius, where shaded radii indicate regions of star-formation activity
  ($I(H\alpha) \ge 30 \text{ R}$ in the narrow component). The broad emission
  is characteristic of an eDIG layer away from star-forming regions, where
  $I(H\alpha)$ is faintest and [NII]$\lambda 6583$/H$\alpha$ is
  highest. Likewise, the broad emission is suggestive of planar gas or of the
  base of the eDIG layer near star-formation activity. Here, $I(H\alpha)$ is
  brighter by factors that range from a few to several orders of magnitude,
  and [NII]$\lambda 6583$/H$\alpha$ is intermediate between values observed in
  HII regions and in diffuse gas. Arrows indicate radii with $I(H\alpha)$
  values that exceed the figure range.}
\end{figure*}

\section{Mass of eDIG Layer}\label{sec_mass}

We estimate the mass of the eDIG layer using the H$\alpha$ surface brightness
to assess the relative importance of the various phases of the gaseous
halo. The H$\alpha$ surface brightness is related to the electron density by:
\begin{equation}
I(H\alpha) = \frac{\int \phi n_{e}^{2} \text{d}l}{2.75 T_{4}^{0.9}},
\end{equation}
where $\phi$ is the volume filling factor and $\int \text{d}l = L$ is the
pathlength through the gas. Here, $I(H\alpha)$ is in Rayleighs, and
$\text{d}l$ is in parsecs. We have corrected $I(H\alpha)$ for inclination
assuming an optically thin disk, and thus the line of sight is taken to be
perpendicular to the disk.

We estimate the characteristic surface brightness of the most diffuse eDIG
detected, $I(H\alpha) = 5.6$ R, by taking the median surface brightness at
$|R| \ge 1$ kpc in slit 1 (see Figure 7). Assuming $I(H\alpha) = 5.6$ R,
$T_{4} = 1$, $L = 1$ kpc, and no variation in the physical conditions along
the line of sight, we find $n_{e} = 0.1 \text{ cm}^{-3}$ and $n_{e} = 0.4
\text{ cm}^{-3}$ for $\phi = 1$ and $\phi = 0.1$, respectively. Our choice of
$L = 1$ kpc follows from the characteristic scale height of the eDIG layer in
the Milky Way and nearby edge-on disk galaxies \citep[e.g.,][]{Rand1997,
  Haffner1999, Collins2001}.

If the eDIG layer extends no farther than the slits ($|R| = 6$ kpc), and is a
uniform disk of height $L = 1$ kpc, then the total mass in the eDIG is
$M_{eDIG} = 8 \times 10^{8} \text{ M}_{\sun}$ ($M_{eDIG} = 3 \times 10^{8}
\text{ M}_{\sun}$) for $\phi = 1$ ($\phi = 0.1$).  This likely underestimates
the total mass, as increased values of $I(H\alpha)$ suggest higher values of
$n_{e}$ near star-forming regions. Although this is a rough estimate subject
to assumptions about the geometry of the layer, the volume filling factor, and
the variation in physical conditions along the line of sight, it is consistent
with estimates of the eDIG mass in other galaxies
\citep[e.g.,][]{Dettmar1990}.

\section{A Dynamical Equilibrium Model}\label{sec_eq}

We now turn to the second goal of this work, to test a dynamical equilibrium
model of the eDIG layer in M83. We ask whether there is sufficient support
available in thermal and turbulent pressure gradients to produce a scale
height characteristic of these layers ($h_{z} = 1$ kpc). Although, critically,
additional support may be found in magnetic field and cosmic ray pressure
gradients (e.g., B16), we lack information about these gradients in face-on
galaxies, and thus do not consider them quantitatively here.

Our dynamical equilibrium model requires that force balance is satisfied in
the vertical and radial directions in an axisymmetic disk:
\begin{equation}
\frac{\partial P(z, R)}{\partial z} = - \rho(z, R) \frac{\partial \Phi(z,
  R)}{\partial z},
\end{equation}
\begin{equation}
  \frac{\partial P(z, R)}{\partial R} = \rho(z, R) \frac{v_{\phi}(z,
    R)^{2}}{R} - \rho(z, R) \frac{\partial \Phi(z, R)}{\partial R}.
\end{equation}
Here, $\frac{\partial \Phi}{\partial z} = g_{z}$ and $\frac{\partial
  \Phi}{\partial R} = g_{R}$ are the gravitational accelerations in the $z$
and $R$ directions, respectively. We construct a mass model of M83 and
determine the galactic gravitational potential in Appendix B. $P = P_{th} +
P_{turb}$ is the sum of the thermal and turbulent pressures, $\rho$ is the gas
density, and $v_{\phi}$ is the azimuthal velocity. Note that we do not include
magnetic or cosmic ray pressure, magnetic tension, or viscosity in our
analysis.

We assume an equation of state of the form
\begin{equation}
P(z, R) = \sigma^{2}(z) \rho(z, R),
\end{equation}
where $\sigma^{2} = \sigma_{th}^{2} + \sigma_{turb}^{2}$ is the quadrature sum
of the thermal and turbulent velocity dispersions. Here, the turbulent
velocity dispersion refers broadly to random motions, and not to a specific
description of turbulence. The observed velocity dispersion shows only local
variations with $R$, and thus we considered only variations in $z$ below.

\begin{figure}[h]
\epsscale{1.2}\plotone{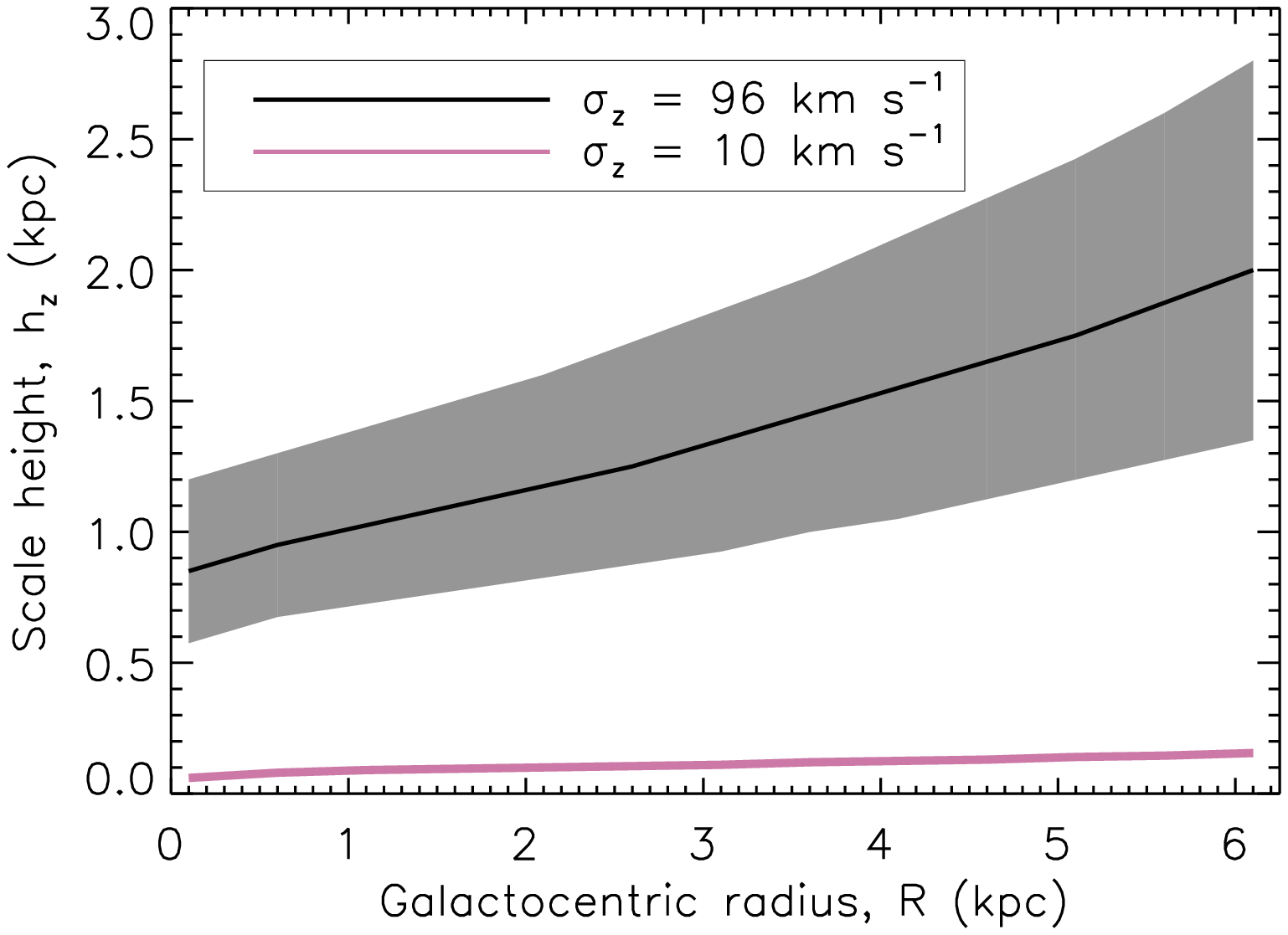}
\caption{The scale height, $h_{z}$, of an isothermal eDIG layer in the
  galactic potential of M83. The black and pink lines show $h_{z}$ for $\sigma
  = 96$ \kms and $\sigma = 10$ \kms, the median observed velocity dispersion
  and the sound speed in the eDIG, respectively. The shaded region shows the
  spread in $h_{z}$ that corresponds to the standard deviation in the observed
  velocity dispersion ($\sigma_{\sigma} = 22$ \kms). If the observed velocity
  dispersion is indicative of a turbulent, cloud-cloud dispersion, then there
  is sufficient support in random motions to produce a characteristic eDIG
  scale height of $h_{z} \ge 1$ kpc.}
\end{figure}

Using the equation of state given in Equation (8), we solve Equation (6) to
determine a general solution for the vertical density profile of the eDIG
layer:
\begin{equation}
  \frac{\rho(z, R)}{\rho(0, R)} = \frac{\sigma^{2}(0)}{\sigma^{2}(z)} \text{exp} \bigg\{ - \int_{0}^{z} \bigg( \frac{\text{d}
    z'}{\sigma^{2}(z')} \frac{\partial \Phi(z', R)}{\partial z'} \bigg)
  \bigg\}.
\end{equation}
If $\sigma$ is independent of $z$, we find the simplified solution below:
\begin{equation}
\rho(z, R) = \rho(0, R) \text{exp}\{-\sigma^{-2}(\Phi(z, R) - \Phi(0, R))\}.
\end{equation}
We define the scale height of the eDIG layer, $h_{z}$, as the value of $z$ at
which $\rho(h_{z})/\rho(0) = \text{e}^{-1}$. Thus, the scale height satisfies
the condition:
\begin{equation}
\sigma^{2} = \Phi(h_{z}, R) - \Phi(0, R).
\end{equation}

In Figure 8, we show the scale height of an eDIG layer with a range of
velocity dispersions in the galactic potential of M83. The scale height of a
layer with the median observed $\sigma = 96$ \kms reaches $h_{z} = 1$ kpc by
$R = 1$ kpc. This is in contrast to the scale height produced by the sound
speed in a $T \sim 10^{4}$ K gas ($\sigma \sim 10$ \kms), as the thermal scale
height reaches only $h_{z} = 0.15$ kpc within $R = 6$ kpc. Thus, if the median
observed $\sigma = 96$ \kms is characteristic of a constant, cloud-cloud
velocity dispersion throughout the eDIG layer, then there is sufficient
support in thermal and turbulent motions to produce a scale height of $h_{z}
\ge 1$ kpc.

We now consider the implications for the azimuthal velocity, $v_{\phi}$, of an
eDIG layer with a constant velocity dispersion. By taking the partial
derivatives of Equations (6) and (7) with respect to $R$ and $z$,
respectively, subtracting the latter from the former, and re-expressing
partial derivatives of $P$ in terms of partial derivatives of $\Phi$, we find
that
\begin{equation}
0 = \frac{-2v_{\phi}\rho(z, R)}{R}\frac{\partial v_{\phi}}{\partial z}.
\end{equation}
Thus, for $\sigma$ constant with $z$, we find $\frac{\partial
  v_{\phi}}{\partial z} = 0$. This model requires that there is no rotational
velocity lag with respect to the disk, a consequence of the Taylor-Proudman
theorem \citep[e.g.,][]{Shore1992}. For a polytropic equation of state, the
Taylor-Proudman theorem states that there is no variation in the gas motions
on a vertical column around the rotational axis. Thus, the gas motions are
defined in the galactic disk, and there is no variation with height above the
disk. The observation that the radial velocity of the eDIG emission tends
toward systemic velocity is not consistent with this class of models.

In Appendix C, we allow $\sigma$ to vary with $z$, and we solve for the
$\sigma(z)$ that satisfies the observed rotational velocity lag,
$\frac{\partial v_{\phi}}{\partial z}$. We find, in summary, that an increase
in $\sigma$ as a function of $z$ is required to reproduce $\frac{\partial
  v_{\phi}}{\partial z}$, but the magnitude of the increase is highly
sensitive to $\epsilon = \sqrt{R\frac{\partial \Phi(h_{HI})}{\partial R}} -
v_{\phi}(h_{HI})$, the difference between the circular velocity and the eDIG
velocity at the HI scale height. Due to the need to fine-tune this dynamical
equilibrium model, and the evidence for local bulk flows near star-forming
regions, we favor a quasi- or non-equilibrium model.

\section{Discussion}\label{sec_disc}

Here, we discuss our results in the context of multi-wavelength observations
of M83 and similar systems, and compare our observations with the predictions
of dynamical equilibrium and non-equilibrium models.

\subsection{A Multi-Phase Gaseous Halo}\label{sec_multi}

There is observational evidence for a multi-phase, gaseous halo in M83 (see
\S1). There are similarities in the extraplanar HI and eDIG properties. Both
phases have rotational velocity lags of a few tens of \kms in projection;
however, the velocity dispersions differ by almost an order of
magnitude. Thus, it is difficult to characterize the relationship between the
diffuse and neutral phases at the disk-halo interface. A model in which the
two phases are directly related - for example, in which the eDIG forms a
``skin'' on condensing, neutral clouds - predicts a comparable velocity
dispersion for the warm and neutral phases. However, as the neutral velocity
dispersion may be underestimated, we cannot rule out such a model. Our eDIG
mass estimate exceeds the extraplanar HI mass by an order of magnitude.  This
suggests that the former is the dominant phase relative to the latter, or that
the former has a very small volume filling factor.

Perhaps the more revealing comparison is between the kinematics of the hot
halo and the eDIG layer. The velocity dispersion of the latter is consistent
with the sound speed in the former, suggesting that the velocity dispersion is
inherited from the hot phase, possibly via entrainment or condensing
clouds. Additionally, the eDIG emission blueshifted with respect to the disk
within $R = 1$ kpc may be produced by warm gas entrained in a hot
outflow. However, due to the kinematics of the bar, the eDIG and disk
velocities are difficult to characterize in this region. The LINER-like
emission-line ratios observed near the nucleus in this work,
\citet{Calzetti2004}, and \citet{Hong2011} may be produced by a nuclear
outflow shocking the surrounding, planar DIG layer.

\subsection{Comparison to Other Galaxies}\label{sec_comp}

We can compare the eDIG properties and kinematics in M83 with that in the
Milky Way and nearby, edge-on disk galaxies. The faintest emission detected in
M83 is an order of magnitude brighter than that observed in the Milky Way by
the Wisconsin H-Alpha Mapper survey \citep[e.g.,][]{Haffner1999,
  Haffner2003}. Thus, we are not sensitive to the most diffuse component of
the eDIG in M83, as the dependence of the H$\alpha$ intensity on the square of
the electron density biases us toward the densest gas along the line of
sight. This is an important consideration as we compare eDIG properties across
galaxies with a range of inclination angles.

In nearby, edge-on disk galaxies, rotational velocity lags of a few tens of
\kms kpc$^{-1}$ are observed in eDIG layers \citep[e.g.,][]{Heald2006a,
  Heald2006b, Heald2007, Fraternali2004, Bizyaev2017}. Assuming that we are
detecting the densest eDIG closest to the disk, the observed median velocity
lag of $\Delta v = -70$ \kms deprojected is steeper than the $\Delta v \sim
-15 - -25$ \kms per scale height observed in other systems
\citep{Heald2007}. However, without a measurement of the scale height in M83,
it is difficult to make a strong statement about the steepness of the
rotational velocity gradient.

The more striking comparison is the velocity dispersion of the eDIG layers. In
the Milky Way, a velocity dispersion of $\sigma(H\alpha) \sim 12$ \kms is
observed toward the north Galactic pole (L. M. Haffner, private
communication). In NGC 891, $\sigma(H\alpha) = 27$ \kms above $z = 1$ kpc
(B16), and several similar systems have $\sigma(H\alpha) = 40 - 60$ \kms,
albeit at lower spectral resolution \citep{Heald2006a, Heald2006b,
  Heald2007}. \citet{Fraternali2004} detect locally broadened H$\alpha$
emission in the moderately-inclined galaxy NGC 2403, with $\sigma(H\alpha) \le
300$ \kms. However, the eDIG layer in M83 is an outlier, as we detect
$\sigma(H\alpha) \sim 100$ \kms all along the slit, with little evidence for a
dependence on underlying disk features.

There are several possible explanations for this discrepancy. First, we may
not be measuring comparable quantities in all galaxies. In an edge-on galaxy,
we may be quantifying $\sigma$ at higher $z$ than in a face-on galaxy; in the
latter case, we can detect the densest eDIG closest to the disk, and the
dependence of the H$\alpha$ intensity on the square of the electron density
biases us toward this denser gas. Additionally, the velocity dispersion in
eDIG layers may be anisotropic, resulting in discrepancies in the dispersions
observed at a range of inclination angles. If so, this would be important for
understanding eDIG dynamics, as the dispersions parallel to the disks are
often assumed to be indicative of those perpendicular to the disks. It is also
possible that the velocity dispersion scales with a galaxy property such as
the star-formation rate. In future work, we will examine the relationship
between vertical velocity dispersion and star-formation rate in face-on disk
galaxies with a range of star-formation properties (Boettcher, Gallagher, \&
Zweibel, in preparation).

\subsection{Comparison of Models}\label{sec_comp}

Here, we consider several models for the dynamical state of the eDIG layer in
M83, and we discuss whether each is consistent with the observations. We pay
particular attention to whether each model can explain the anomalously large
$\sigma$ observed in this system.

\textit{A Dynamical Equilibrium Model}: First, we consider the dynamical
equilibrium model that we tested in \S6. We found that we are able to
reproduce the characteristic scale heights of eDIG layers ($h_{z} \sim 1$ kpc)
from the observed velocity dispersion ($\sigma = 96$ \kms). However,
maintaining a reasonable energy requirement while reproducing the observed
rotational velocity gradient requires fine-tuning of the model.

From our mass estimate in \S5, the kinetic energy in random motions for
$\sigma = 96$ \kms is on the order of $KE \sim 10^{56}$ ergs. Following
\cite{Draine2011}, the cooling time for a shock with $v_{s} \sim 100$ \kms in
a medium with $n_{e} \sim 0.1$ cm$^{-3}$ is $t_{cool} \sim 7 \times 10^{4}$
years. Thus, the cloud collision timescale must be long if the energy
requirement is to remain reasonable. The likelihood of eDIG cloud collisions
is reduced if this phase has a very small volume filling factor and is
embedded in a hot halo. This suggests a picture in which the warm phase is
condensing out of or evaporating into the hot phase, with a velocity
dispersion that is characteristic of the sound speed in the hotter medium.

\textit{A Galactic Fountain Model}: A galactic fountain flow describes the
circulation of gas between the disk and the halo due to star-formation
feedback \citep{Shapiro1976, Bregman1980}. It is thought that the gas leaves
the disk in a hot phase and returns to the disk after cooling, passing through
a warm, ionized phase. However, it is not clear during which part of the cycle
a warm ionized phase is present (e.g., whether warm gas is entrained in a hot
outflow, or condenses from the hot phase).

The predictions of galactic fountain models include a rotational velocity
gradient, local outflows from star-forming regions, and launch velocities
around $\sim 100$ \kms \citep{Shapiro1976}; all of these predictions are
consistent with our observations. Additionally, a fountain flow that is
largely in the $z$ direction is consistent with the smaller and larger
$\sigma$ observed in high- and low-inclination galaxies, respectively. In this
model, the large $\sigma$ results from local outflows near the disk as well as
quasi-symmetric inflow and outflow near the turn-around height. The latter
scenario is feasible due to the comparitively long time that a cloud spends at
its maximum height, and is necessary to explain detections away from
star-forming regions as well as the large $\sigma$ and $\partial
v_{\phi}/\partial z$ observed along the same line of sight.

\textit{A Galactic Wind Model}: The high X-ray surface brightness and the
LINER-like emission-line ratios near the nucleus may be evidence of a hot,
ionized outflow from the central kpc of M83. The eDIG within this region is
largely blueshifted with respect to the disk, and may be entrained in an
outflow. There is also evidence of local outflows from star-forming spiral
arms. However, the eDIG layer at large does not appear to be associated with a
galactic wind. The radial velocities are dominated by the rotational velocity
of a lagging halo, and not by a preferentially blueshifted or redshifted
flow. The relationship between the origin and evolution of eDIG layers and
galactic outflows is of interest for further study.

\textit{An Accretion Model}: We disfavor an accretion flow as the origin of
the eDIG layer for several reasons. Firstly, the high
[NII]$\lambda$6583/H$\alpha$ and [SII]$\lambda$6717/H$\alpha$ line ratios
suggest that the gas is chemically enriched. If it is embedded in an accretion
flow, the origin must be the enriched halo and not the pristine intergalactic
medium. Secondly, the eDIG layer follows the rotation curve of the disk,
albeit at a reduced rotational velocity. This is consistent with gas that
originated in the disk, was lifted into the halo, and was radially
redistributed to conserve angular momentum. However, the evidence for
interaction in the extended HI disk of M83 should be kept in mind when
considering the kinematics of the halo.

Thus, both a dynamical equilibrium model and a galactic fountain flow are
broadly consistent with the observations. However, the need to fine-tune the
former model leads us to favor the latter. The true dynamical state may be
somewhere in between these models. Regardless, the importance of the hot (and
potentially the cold) phase is clear, and emphasizes the need for a
multi-wavelength approach to modeling these layers. For example, the mass
hierarchy of hot, warm, and neutral gas suggested by this analysis may imply
an energy flow: explusion of gas from the disk in the hot phase, condensation
into clouds to produce the warm phase, and cloud-cloud collisions to cool to a
neutral phase. The ability to address questions of energy balance and dynamics
simultaneously is of interest for future work.

\section{Summary and Conclusions}\label{sec_conc}

Using optical emission-line spectroscopy from the Robert Stobie Spectrograph
on the Southern African Large Telescope, we performed the first detection and
kinematic study of extraplanar diffuse ionized gas in the nearby, face-on disk
galaxy M83. A Markov Chain Monte Carlo method was used to decompose the
[NII]$\lambda\lambda$6548, 6583, H$\alpha$, and [SII]$\lambda\lambda$6717,
6731 emission lines into contributions from HII region, planar DIG, and
extraplanar DIG emission. The eDIG layer is clearly identified by its
emission-line ratios ([NII]$\lambda$6583/H$\alpha \gtrsim 1.0$), velocity
dispersion ($\sigma = 96$ \kms), and rotational velocity lag with respect to
the disk. The main results are as follows:

\begin{itemize}
\item The median, line-of-sight velocity dispersion observed in the diffuse
  gas, $\sigma = 96$ \kms, is a factor of a few higher than that observed in
  the Milky Way and nearby, edge-on disk galaxies. This suggests that the
  velocity dispersions in these layers may be anisotropic; however, further
  observations of the velocity dispersions in face-on eDIG layers are needed.
\item The diffuse emission lags the disk emission in rotational velocity,
  qualitatively consistent with the multi-phase, lagging halos observed in
  other galaxies. The median velocity lag between the disk and the halo is
  $\Delta v = -24$ \kms in projection, or $\Delta v = -70$ \kms corrected for
  inclination. This exceeds the rotational velocity lags of $\Delta v \sim -15
  - -25$ \kms per scale height observed in several nearby, edge-on disk
  galaxies \citep{Heald2007}.
\item If the velocity dispersion is indicative of turbulent (random) motions,
  there is sufficient thermal and turbulent support to produce an eDIG scale
  height of $h_{z} \sim 1$ kpc in dynamical equilibrium. This model does not
  require vertical support from magnetic field or cosmic ray pressure
  gradients, consistent with a largely vertically-oriented (``X-shaped'')
  field. However, reproducing the observed velocity dispersion and rotational
  velocity gradient while keeping the energy requirement reasonable requires a
  finely tuned model.
\item We favor a quasi- or non-equilibrium model for the eDIG layer. There is
  evidence of local bulk flows near star-forming regions that may trace the
  warm, ionized phase of a galactic fountain flow.
\item Multi-wavelength observations of M83 reveal extraplanar hot and cold
  gas. The velocity dispersion of the eDIG layer is consistent with the sound
  speed in the hot phase, and rotational velocity lags are observed in both
  the cold and warm components. The relationship between the energetics and
  dynamics of these phases is of interest for future study.
\end{itemize}

In future work, we will construct a sample of both face-on and edge-on
galaxies, develop a picture of the three-dimensional kinematics of eDIG
layers, and contextualize this picture in the multi-phase environment of the
disk-halo interface.

\acknowledgments \textit{Acknowledgments:} All of the observations reported in
this paper were obtained with the Southern African Large Telescope (SALT). We
thank the SALT astronomers and telescope operators for obtaining the
observations and Petri Vaisanen for advice on data acquisition. We acknowledge
Bob Benjamin for useful comments and discussion, Ken Nordsieck for his
expertise on the RSS instrument, Arthur Eigenbrot for help with data
reduction, and Eowyn Liu for assistance with emission-line fitting. We thank
Masataka Okabe and Kei Ito for supplying the colorblind-friendly color palette
used in this paper (see \url{fly.iam.u-tokyo.ac.jp/color/index.html}). We
acknowledge helpful comments from the anonymous referee that improved the
clarity of the paper.

This material is based upon work supported by the National Science Foundation
Graduate Research Fellowship Program under Grant No. DGE-1256259. Any
opinions, findings, and conclusions or recommendations expressed in this
material are those of the author(s) and do not necessarily reflect the views
of the National Science Foundation. Support was also provided by the Graduate
School and the Office of the Vice Chancellor for Research and Graduate
Education at the University of Wisconsin-Madison with funding from the
Wisconsin Alumni Research Foundation.

This work has made use of NASA's Astrophysics Data System and of the NASA/IPAC
Extragalactic Database (NED) which is operated by the Jet Propulsion
Laboratory, California Institute of Technology, under contract with the
National Aeronautics and Space Administration.

\bibliographystyle{apj}

\begin{appendices}

\section{Markov Chain Monte Carlo Convergence}

Here, we discuss the convergence of the Markov Chain Monte Carlo method used
in this work. We use $N = 10^{5}$ links in the MCMC chain, and reject the
first $N = 10^{4}$ links as the ``burn-in'' period. In Figure 9, we
demonstrate the convergence of the method at $N = 10^{5}$ using all spectra
from slit 1 with multi-component fits. For $N = 1.25 \times 10^{4}, N = 2
\times 10^{4}, N = 5 \times 10^{4}, \text{ and } N = 2 \times 10^{5}$, we
compare the normalized difference between the best-fit parameters at each $N$
and at $N = 10^{5}$ ($\Delta P = |P(N) - P(N = 10^{5})|/P(N = 10^{5})$) with
the normalized uncertainties on the best-fit parameters at $N = 10^{5}$
($\sigma_{P}(N = 10^{5})/P(N = 10^{5})$). We can conclude that the method has
converged within the errors at $N = 10^{5}$ if the former quantity is smaller
than the latter at larger $N$. For $N = 1.25 \times 10^{4}$ and $N = 2 \times
10^{4}$, we find that $\Delta P > \sigma_{P}(N = 10^{5})/P(N = 10^{5})$ for
22\% and 10\% of the best-fit parameters, respectively. In comparison, for $N
= 5 \times 10^{4}$ and $N = 2 \times 10^{5}$, this is reduced to an acceptable
level of scatter at only 2\%. Thus, the best-fit parameter values are largely
unchanged within the uncertainties beyond $N = 5 \times 10^{4}$, and we
conclude that the MCMC method is converged at our choice of $N = 10^{5}$.

\begin{figure}[h]
\epsscale{1.0}\plotone{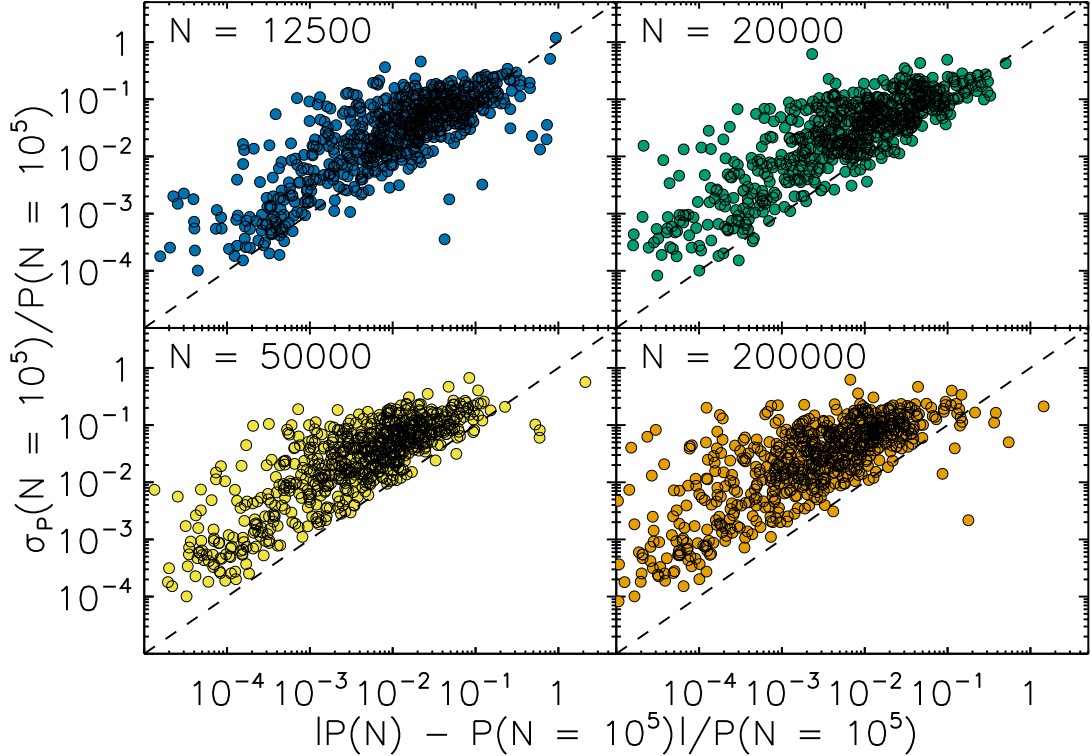}
\caption{The normalized difference between the best-fit parameters at $N$ and
  at $N = 10^{5}$ links in the MCMC chain, $|P(N) - P(N = 10^{5})|/P(N =
  10^{5})$ as compared to the normalized parameter uncertainties at $N =
  10^{5}$ ($\sigma_{P}(N = 10^{5})/P(N = 10^{5})$) for a range of $N$. Each
  point represents a best-fit parameter value for a spectrum in slit 1 with a
  multi-component fit. The dashed line denotes equality. For $N = 5 \times
  10^{4}$ and $N = 2 \times 10^{5}$, only 2\% of the best-fit parameter values
  vary with respect to $N = 10^{5}$ by more than their uncertainties at $N =
  10^{5}$, suggesting that the MCMC method is converged within the errors for
  $N > 5 \times 10^{4}$. For all $N$, we reject the first $N = 10^{4}$ links
  as the ``burn-in'' period.}
\end{figure}

\section{A Mass Model for M83}

Here, we construct a mass model for M83 to determine the galactic
gravitational potential. \citet{Herrmann2009} develop a mass model of the disk
of M83 using the observed vertical velocity dispersion of planetary nebulae
($\sigma_{z,PNe}$) over a wide range of galactocentric radii ($R \le$ 6
$R$-band scalelengths). The authors use a thin and thick disk model to
reproduce the relatively flat distribution of $\sigma_{z,PNe}$ at large
$R$. For both disks, they assume a vertical density distribution of the form:
\begin{equation}
\rho(z) = \rho_{0}\text{sech}^{2/n}\bigg(\frac{nz}{2h_{z}}\bigg).
\end{equation} 
Here, $n = 2$, as compared to the isothermal ($n = 1$) and exponential ($n =
\infty$) cases. Assuming exponential disks in $R$, the density distribution is
the sum of the thin ($t$) and thick ($th$) components:
\begin{equation}
\begin{split}
  \rho(z,R) =
  \rho_{0,t}\text{e}^{-R/h_{R,t}}\text{sech}\bigg(\frac{z}{h_{z,t}}\bigg) +
  \rho_{0,th}\text{e}^{-R/h_{R,th}}\text{sech}\bigg(\frac{z}{h_{z,th}}\bigg).
\end{split}
\end{equation} 

The central velocity dispersions, $\sigma_{z}(0)$, vertical scale heights,
$h_{z}$, and radial scale lengths, $h_{R}$, of these disk components are given
in Table 3. For each component, we calculate central mass volume densities,
$\rho_{0}$, from the central mass surface density, $\Sigma_{0}$, for $n = 2$
disks: $\sigma_{z}^{2}(R) = 1.7051\pi G\Sigma(R)h_{z}$. The total mass in the
thin and thick disks is $5.8 \times 10^{10} \Msun$ and $3.6 \times 10^{10}
\Msun$, respectively. Stellar mass estimates from \textit{Wide-field Infrared
  Survey Explorer} (\textit{WISE}) data suggest that this exceeds the baryonic
mass by a factor of a few \citep{Jarrett2013}. This is a consequence of the
low mass-to-light ratio required by \citet{Herrmann2009} to reproduce the
relatively flat distribution in $\sigma_{z,PNe}$ as a function of $R$. Our
goal is to quantify whether thermal and turbulent motions can support the eDIG
layer at $h_{z} = 1$ kpc. Thus, we favor an over-massive disk rather than an
under-massive one, so that we may be sure that a successful model is not a
result of underestimating the mass in the disk.

To reproduce the rotation curve, we add a dark matter halo with a
Navarro-Frenk-White (NFW) profile of the form:
\begin{equation}
  \rho_{DM}(R) = \frac{\rho_{0,DM}}{R/a_{DM}(1 + R/a_{DM})^{2}},
\end{equation}
where $\rho_{0,DM}$ is the central dark matter density and $a_{DM}$ is the
scale radius. A range of inclination angles, $i$, and position angles, $PA$,
have been suggested for M83, with evidence that the former increases and the
latter decreases with $R$ \citep[e.g.,][]{Huchtmeier1981, Heald2016}. We
choose $i = 24^{\circ}$ and $PA = 226^{\circ}$, the smallest inclination
determined for the inner disk, to once again favor the most massive
model. With these assumptions, the HI rotation curve has a maximum velocity of
$v_{c,max} \sim 255$ \kms between $R \sim 10 - 20$ kpc \citep{Park2001,
  Heald2016}.

We determine the values of $\rho_{0,DM}$ and $a_{DM}$ as follows. We test
values of $a_{DM}$ between $1 \text{ kpc} \le a_{DM} \le 30 \text{ kpc}$; for
each value of $a_{DM}$, we solve for the value of $\rho_{0,DM}$ that most
closely produces $v_{c,max} = 255$ \kms between $10 \text{ kpc} \le R \le 20
\text{ kpc}$. We then quantify the quality of the fit in a least-squares sense
using our optical rotation curve for $R \le 6$ kpc. The quality of the fit
increases with increasing $a_{DM}$ at first, and then becomes fairly flat with
increasing $a_{DM}$ beyond $a_{DM} = 20$ kpc. Thus, we choose $a_{DM} = 20$
kpc and $\rho_{0,DM} = 6 \times 10^{6} \Msun \text{kpc}^{-3}$. We exclude
velocities at $R \le 0.5 \text{ kpc}$ from this analysis due to the influence
of the bar.

The gravitational potential of the thin and thick disks is of the form:
\begin{equation}
\begin{split}
  \Phi(z,R) = -\frac{4G\Sigma_{0}}{h_{R}} \int_{-\infty}^{\infty}
  \mathrm{d}z^{\prime} \, \Big[\text{sech}\bigg(\frac{z}{h_{z}}\bigg)
  \int_{0}^{\infty} \mathrm{d}a \, \arcsin \Big(\frac{2a}{S_{+} + S_{-}}
  \Big)aK_{0} \Big(\frac{a}{h_{R}} \Big) \Big],
\end{split}
\end{equation}     
where $S_{\pm} \equiv \sqrt{(z - z^{\prime})^{2} + (a \pm R)^{2}}$,
$\Sigma_{0}$ is the central mass surface density, and $K_{0}$ is the zeroth
order modified Bessel function \citep{Cuddeford1993}. The equivalent
expression for the dark matter halo is given by:
\begin{equation}
  \Phi(z,R)_{DM} = -4 \pi G\rho_{0,DM}a_{DM}^{2}\frac{\text{ln}(1 + R/a_{DM})}{R/a_{DM}}.
\end{equation}
From the gravitational potential of the thin disk, thick disk, and dark matter
halo, we calculate a model rotation curve, $v_{c} = \sqrt{R\frac{\partial
    \Phi}{\partial R}}$, and compare to the observed rotation curve in Figure
10.

\begin{figure}[h]
\epsscale{1.0}\plotone{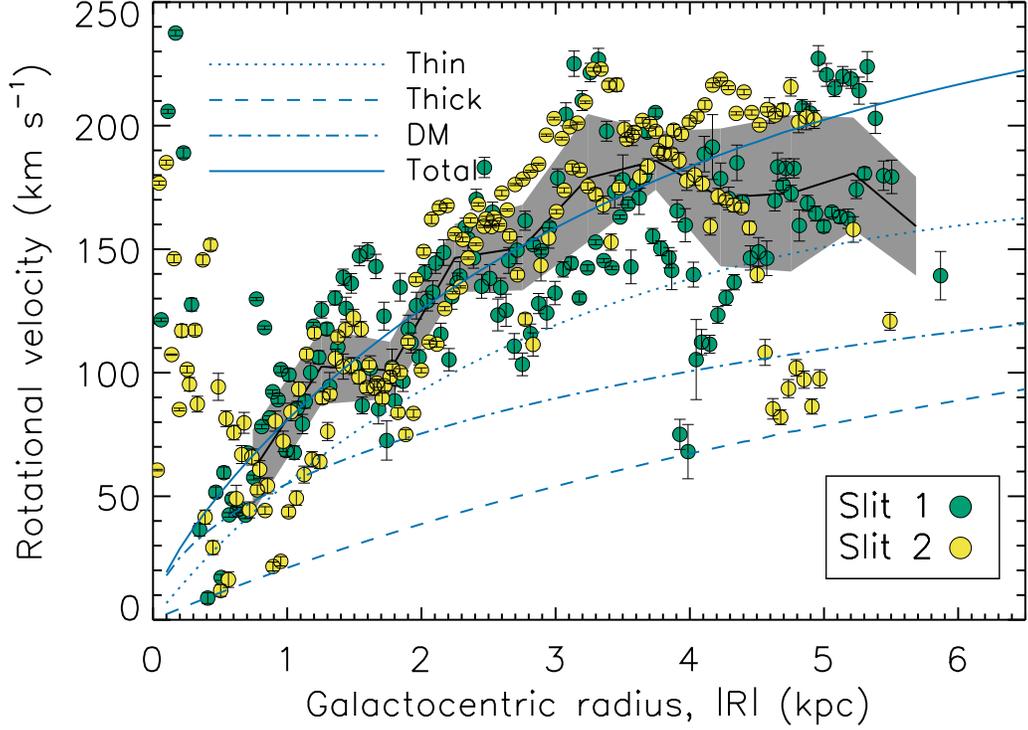}
\caption{The observed and modeled rotation curves of M83. The observed
  velocities from slits 1 (green points) and 2 (yellow points) are shown
  corrected for inclination and position angle ($v_{\phi} = (v_{obs} -
  v_{sys})/(\text{cos}(\phi)\text{sin}(i))$). For observations on the
  southwest side of the galaxy, the signs of $R$ and $v_{\phi}$ are changed to
  allow overplotting. The broad component is not included in this
  analysis. The black line and shaded region indicate the median and median
  absolute deviation of the observed rotation curve in bins of $\Delta R =
  0.5$ kpc, respectively. Significant scatter around the median is expected
  due to velocities in the $R$ and $z$ directions as well as spiral
  structure. Velocities within $|R| = 1$ kpc are highly affected by the bar
  kinematics. The rotation curve of the modeled thin disk (dotted blue line),
  thick disk (dashed blue line), and dark matter halo (dot-dashed blue line)
  are overplotted, with the quadrature sum of these components shown as the
  solid blue line.}
\end{figure}

\begin{deluxetable}{ccc}[t]
\tabletypesize{\scriptsize}
\tablecolumns{4}
\tablewidth{0pt}
\tablecaption{M83 Mass Model}
\tablehead{ 
\colhead{Parameter} &
\colhead{Value} &
\colhead{Reference \tablenotemark{a}}
}
\startdata
$\sigma_{z,t}(0)$ & 73 \kms & (1) \\
$\rho_{0,t}$ & $5 \times 10^{8} \Msun \text{kpc}^{-3}$ & (1) \\
$h_{z,t}$ & 0.4 kpc & (1) \\
$h_{R,t}$ & 4 kpc & (1) \\
$\sigma_{z,th}(0)$ & 40 \kms & (1) \\
$\rho_{0,th}$ & $2 \times 10^{7} \Msun \text{kpc}^{-3}$ & (1) \\
$h_{z,th}$ & 1.2 kpc & (1) \\
$h_{R,th}$ & 10 kpc & (1) \\
$a_{DM}$ & 20 kpc & (2) \\
$\rho_{0,DM}$ & $6 \times 10^{6} \Msun \text{kpc}^{-3}$ & (2)
\enddata
\tablenotetext{a}{References: (1) \citet{Herrmann2009}; (2) this work.}
\end{deluxetable}

\section{A Dynamical Equilibrium Model: Implications for $\sigma(z)$}

Here, we consider a dynamical equilibrium model for the eDIG layer in M83 in
which $\sigma$ is allowed to vary with $z$. To solve for the $\sigma(z)$ that
satisfies the observed rotational velocity lag, $\frac{\partial
  v_{\phi}}{\partial z}$, we take partial derivatives of Equations (6) and (7)
with respect to $R$ and $z$, respectively. We subtract one from the other,
assuming that $\sigma$ varies much more rapidly in $z$ than in $R$
($\frac{\partial \sigma^{-2}}{\partial R}$ = 0). This yields:
\begin{equation}
  0 = \frac{\partial \sigma^{-2}}{\partial z}\bigg(\frac{\partial
    \Phi}{\partial R} - \frac{v_{\phi}^{2}}{R}\bigg) -
  \frac{\sigma^{-2}}{R}\frac{\partial v_{\phi}^{2}}{\partial z}. 
\end{equation}

Integrating Equation (18) with respect to $z$, we find:
\begin{equation}
  \sigma(z_{2}) = \sigma(z_{1}) \bigg[ \text{exp} \bigg\{ \int_{z_{1}}^{z_{2}} \frac{1}{R}
  \frac{\partial v_{\phi}^{2}}{\partial z'} \bigg( \frac{\partial \Phi}{\partial
    R} - \frac{v_{\phi}^{2}}{R} \bigg)^{-1} \text{d}z' \bigg\} \bigg]^{-1/2}.
\end{equation}
We evaluate this expression as follows. We assume that the gaseous disk is
co-rotating within the HI thin disk scale height, $z < h_{HI} = 0.1 \text{
  kpc}$. At $z = h_{HI}$, we set $v_{\phi}(h_{HI}) = \sqrt{R\frac{\partial
    \Phi(h_{HI})}{\partial R}} - \epsilon$. At $z > h_{HI}$, we impose a
rotational velocity gradient such that $v_{\phi}(z) = v_{\phi}(h_{HI}) +
\frac{\partial v_{\phi}}{\partial z}z$. We do not allow the gas to
counter-rotate.

The use of an $\epsilon$ term is necessary to avoid the divergence of the
integral in Equation (19) at $z = h_{HI}$, and follows physically from the
reduction of the rotational velocity with respect to the circular velocity due
to an outward pressure gradient. A preferred value of $\epsilon$ is found by
setting $\frac{\partial P}{\partial R} = \frac{\sigma^{2}\rho}{h_{R,t}}$ in
Equation (7), where $h_{R,t}$ is the thin disk radial scale length. At $R =
h_{R,t}$, $\epsilon = 30$ \kms; since $\sigma(z)$ is highly sensitive to
$\epsilon$, and we consider both a small ($\epsilon = 3$ \kms) and a preferred
($\epsilon = 30$ \kms) value.

The observations favor a fairly steep rotational velocity gradient. The median
observed difference in radial velocity between the narrow and the broad
component is $\Delta v = -24$ \kms in projection, or $\Delta v = -70$ \kms
corrected for inclination. If we are detecting gas at the scale height, $h_{z}
\ge 1$ kpc, then the rotational velocity lag is at least a few tens of \kms
kpc$^{-1}$. If, as expected, we are detecting the densest eDIG closest to the
disk, $z < h_{z}$, then the rotational velocity lag may be even steeper.

\begin{figure}[h]
\epsscale{1.0}\plotone{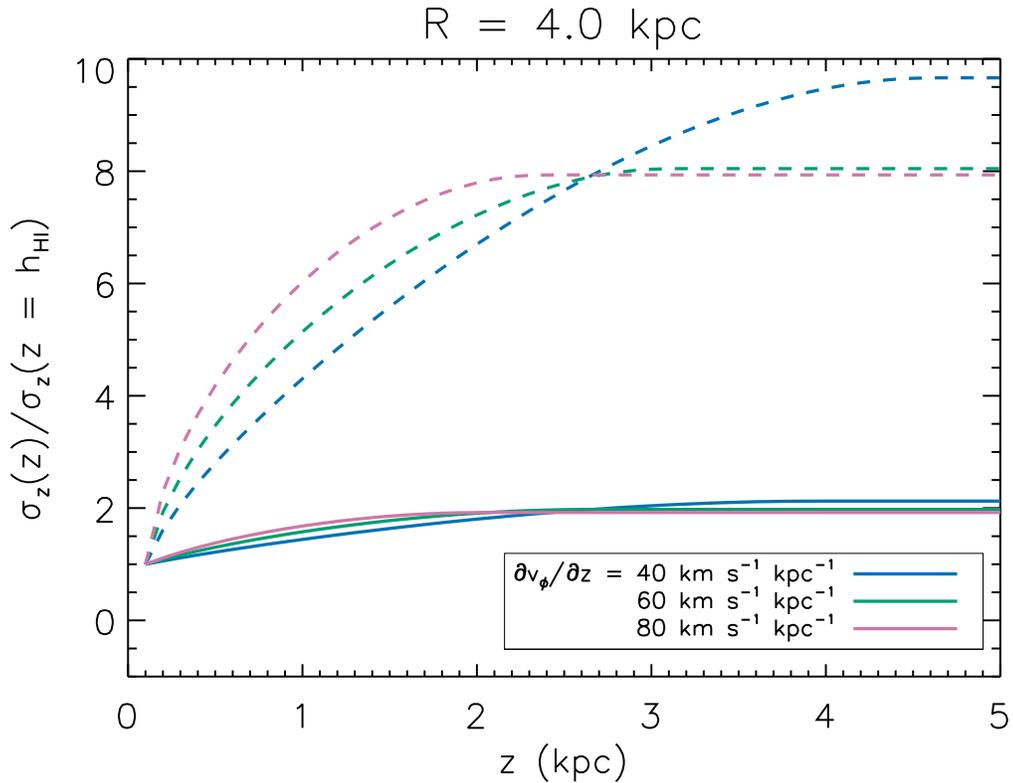}
\caption{$\sigma(z)$ required to produce a range of rotational velocity
  gradients in dynamical equilibrium for $\epsilon = \sqrt{R\frac{\partial
      \Phi(h_{HI})}{\partial R}} - v_{\phi}(h_{HI}) = 3$ \kms (dashed lines)
  and $\epsilon = 30$ \kms (solid lines). For the preferred (larger)
  $\epsilon$, $\sigma$ must increase by a factor of two within $z = 5$
  kpc. However, for the smaller $\epsilon$, $\sigma$ must increase by an order
  of magnitude, posing a problem for the energetics of the system. The result
  is shown at $R = 4$ kpc, but similar results are found at other
  galactocentric radii. Without knowledge of $\epsilon$, this class of models
  is highly unconstrained, and we regard the solutions as contrived.}
\end{figure}

In Figure 11, we show the $\sigma(z)$ required to produce a range of
$\frac{\partial v_{\phi}}{\partial z}$ in dynamical equilibrium at $R = 4$
kpc. Similar results are found at other galactocentric radii. Without
knowledge of $\epsilon$, this class of models is highly unconstrained. For
$\epsilon = 30$ \kms (solid lines), an increase in $\sigma$ by a factor of two
is required to produce a rotational velocity gradient of at least a few tens
of \kms kpc$^{-1}$. For $\epsilon = 3$ \kms (dashed lines), an increase in
$\sigma$ by an order of magnitude is required instead. Taken at face value,
this suggests a velocity dispersion of several hundred \kms at large $z$,
consistent with the sound speed in a $T = 10^{7}$ K gas. As this is an order
of magnitude hotter than expected for the hot halo gas, small values of
$\epsilon$ present a problem for the energetics of the system.

Thus, simultaneously satisfying the observed velocity dispersion and
rotational velocity gradient while keeping the energy requirements reasonable
requires fine-tuning of the model. We regard this model as contrived, but
present it for completeness. This result emphasizes the importance of deep
spectroscopic observations of edge-on eDIG layers, in which $\sigma(z)$ and
$\frac{\partial v_{\phi}}{\partial z}$ can be quantified and this class of
model can be constrained.

\end{appendices}
\end{document}